%% file: ms.tex
\documentclass[12pt,preprint]{aastex}
\usepackage{natbib,psfig}

\newcommand{\Msun}{\mbox{\,$\rm{M_{\odot}}$}} 
\newcommand{\Rsun}{\mbox{\,$\rm{R_{\odot}}$}} 
\newcommand{\Lsun}{\mbox{\,$\rm{L_{\odot}}$}} 

\newcommand{\Xsun}{\mbox{\,$\rm{X_{\odot}}$}}
\newcommand{\Tstar}{\mbox{\,$T_{*}$}}
\newcommand{\Teff}{\mbox{\,$T_{eff}$}}
\newcommand{\Rstar}{\mbox{\,$R_{*}$}} 
\newcommand{\Reff}{\mbox{\,$R_{2/3}$}} 
\newcommand{\Rtrans}{\mbox{\,$R_{t}$}}

\newcommand{\Mdot}{\mbox{\,$\dot{M}$}}

\newcommand{\Mdotc}{\mbox{\,$\dot{M}_{C}$}}
\newcommand{\Mdots}{\mbox{\,$\dot{M}_{S}$}}

\newcommand{\vturb}{\mbox{\,$v_{turb}$}}
\newcommand{\vedge}{\mbox{\,v$_{edge}$}}
\newcommand{\vinf}{\mbox{\,v$_{\infty}$}}
\newcommand{\logg}{\mbox{\,$\log{g}$}}
\newcommand{\logN}{\mbox{\,$\log{N}$}}
\newcommand{\E}[1]{\mbox{\,$\rm x 10^{#1}$}}
\newcommand{\XH}{\mbox{\,$X_{H}$}}
\newcommand{\XHe}{\mbox{\,$X_{He}$}}
\newcommand{\XC}{\mbox{\,$X_{C}$}}
\newcommand{\XN}{\mbox{\,$X_{N}$}}
\newcommand{\XO}{\mbox{\,$X_{O}$}}
\newcommand{\XSi}{\mbox{\,$X_{Si}$}}
\newcommand{\XS}{\mbox{\,$X_{S}$}}
\newcommand{\XP}{\mbox{\,$X_{P}$}}
\newcommand{\XFe}{\mbox{\,$X_{Fe}$}}

\newcommand{\Htwo}{\mbox{\rm{H}$_2$}}
\newcommand{\HI}{\mbox{\rm{\ion{H}{1}}}}

\newcommand{\HeI}{\ion{He}{1}}
\newcommand{\HeII}{\ion{He}{2}}

\newcommand{\CIII}{\ion{C}{3}}
\newcommand{\CIV}{\ion{C}{4}}
\newcommand{\NI}{\ion{N}{1}}

\newcommand{\NIII}{\ion{N}{3}}
\newcommand{\NIV}{\ion{N}{4}}
\newcommand{\NV}{\ion{N}{5}}

\newcommand{\OI}{\ion{O}{1}}
\newcommand{\OII}{\ion{O}{2}}
\newcommand{\OIII}{\ion{O}{3}}
\newcommand{\OIV}{\ion{O}{4}}
\newcommand{\OV}{\ion{O}{5}}
\newcommand{\OVI}{\ion{O}{6}}
\newcommand{\NeIII}{\ion{Ne}{3}}
\newcommand{\NeIV}{\ion{Ne}{4}}
\newcommand{\NeV}{\ion{Ne}{5}}

\newcommand{\MgV}{\ion{Mg}{5}}

\newcommand{\SiIII}{\ion{Si}{3}}
\newcommand{\SiIV}{\ion{Si}{4}}

\newcommand{\PV}{\ion{P}{5}}
\newcommand{\SII}{\ion{S}{2}}

\newcommand{\SIV}{\ion{S}{4}}

\newcommand{\SVI}{\ion{S}{6}}

\newcommand{\ArIV}{\ion{Ar}{4}}
\newcommand{\ArV}{\ion{Ar}{5}}

\newcommand{\FeV}{\ion{Fe}{5}}
\newcommand{\FeVI}{\ion{Fe}{6}}
\newcommand{\FeVII}{\ion{Fe}{7}}
\newcommand{\FeVIII}{\ion{Fe}{8}}

\newcommand{\eg}{\emph{e.g.}}
\newcommand{\ie}{\emph{i.e.}}
\newcommand{\doublet}{$\lambda\lambda$}
\newcommand{\singlet}{$\lambda$}
\newcommand{\res}{\mbox{\,$\Delta\lambda$}}

\newcommand{\EBMV}{\mbox{\,$E_{\rm{B-V}}$}}

\newcommand{\tlusty}{TLUSTY}

\newcommand{\Hbeta}{H$\beta$}
\newcommand{\Hgamma}{H$\gamma$}
\newcommand{\Hdelta}{H$\delta$}

\newcommand{\Lya}{Ly$\alpha$}
\newcommand{\Lyb}{Ly$\beta$}
\newcommand{\Lyg}{Ly$\gamma$}

\newcommand{\Zsun}{\mbox{\,$\rm{Z_{\odot}}$}}
\newcommand{\Telec}{\mbox{\,$T_{e}$}}
\newcommand{\nelec}{\mbox{\,$n_{e}$}}
\newcommand{\kms}{\mbox{\,$\rm{km\:s^{-1}}$}}

\newcommand{\Msunyr}{\mbox{\,$\rm{M_{\odot}\:yr^{-1}}$}}
\newcommand{\flam}{\mbox{\,$\rm{erg\:s^{-1}\:cm^{-2}\:\AA^{-1}}$}}

\begin{document}

\title{CENTRAL STARS OF PLANETARY NEBULAE IN THE LARGE MAGELLANIC
  CLOUD: A FAR-UV SPECTROSCOPIC ANALYSIS\footnote{Based on observations made
  with the NASA-CNES-CSA Far Ultraviolet Spectroscopic Explorer and
  archival data. FUSE is operated for NASA by the Johns Hopkins
  University under NASA contract NAS5-32985.}}

\author{J.E. Herald, L. Bianchi}
\vspace{1mm}
\affil{Department of Physics and Astronomy, The Johns Hopkins University}
\authoraddr{3400 N. Charles St., Baltimore, MD 21218-2411}

\begin{abstract}
We observed seven central stars of planetary nebulae (CSPN) in the
Large Magellanic Cloud (LMC) with the \emph{Far Ultraviolet
Spectroscopic Explorer} (FUSE), and performed a model-based
analysis of these spectra in conjunction with
\emph{Hubble Space Telescope} (HST) spectra in the UV and optical
range to determine the stellar and nebular parameters.  Most of the
objects show wind features, and they have effective temperatures
ranging from 38 to 60~kK with mass-loss rates of $\simeq
5\E{-8}$~\Msunyr.  Five of the objects have typical LMC abundances.
One object (SMP LMC~61) is a [WC4] star, and we fit its spectra with
He/C/O-rich abundances typical of the [WC] class, and find its
atmosphere to be iron-deficient.  Most objects have very hot ($T \gtrsim
2000$~K) molecular hydrogen (\Htwo) in their nebulae, which may
indicate a shocked environment.  One of these (SMP LMC~62) also
displays \OVI\ 1032-38 nebular emission lines, rarely observed
in PN.
\end{abstract}


\section{INTRODUCTION}\label{sec:intro}

With respect to the study of planetary nebulae (PN) systems, those in
the Large Magellanic Cloud (LMC) are important for two primary
reasons.  Attempts to describe the evolution of Galactic PN are
hindered by large relative uncertainties in their distances, which
carry over into physical parameters such as the stellar luminosity and
radius of the central star of the PN (CSPN), and the size and ionized
mass of the PN itself.  This obstruction is removed for the PN of the
LMC, all of which lie at essentially at the same distance, allowing
the physical parameters to be scaled to absolute values.
Additionally, the lower metallicity of the LMC relative
to the Milky Way allows the role of metallicity in
low/intermediate-mass stellar evolution to be assessed.  The largest
impact a higher metallicity is expected to have on a star's evolution
is a more efficient radiative driving during its windy phases: 
the asymptotic giant branch (AGB) phase, the post-AGB phase, and/or
perhaps a Wolf-Rayet phase [WR].  This would increase the star's
mass-loss rate, and thus slow its
evolution during these periods.  Such an effect would manifest itself
in the form of different relative population ratios in galaxies of
varying metallicities.  These implications carry over into galaxy
evolution, through chemical evolution and the dynamic interactions
between the star's ejected material and the surrounding interstellar
medium (ISM).  Although these effects are not as dramatic on an
individual basis as those of a massive star, the large fraction of
stars that evolve through the CSPN phase make their contribution to the
Galactic chemical evolution significant (see, \eg, \citealp{marigo:01} for a
discussion).

Characterization of an individual PN system requires an understanding
of both the PN and its central star.  Several non-LTE (NLTE)
spectroscopic studies of Galactic CSPN in the optical range have been
carried out, both of CSPN without winds using plane-parallel analyses
(\eg, \citealp{mendez:85,herrero:90}) and of CSPN with winds using
spherical codes (\eg,
\citealp{leuenhagen:98,koesterke:97,demarco:98}).  The sample of known
LMC PN for which HST spectroscopy exists includes only the brightest
and rather compact objects.  The nebular continuum of compact (high
density) PN typically masks the light of the central star for
wavelengths longwards of $\sim$1215~\AA\ \citep{bianchi:97},
complicating the task of characterizing the two components if relying
solely on UV and optical data.  Characteristics of a large sample of
LMC CSPN have been inferred from photoionization models of their
nebular spectra (\eg,
\citealp{dopita:91a,vassiliadis:96s,vassiliadis:98s}).  A handful have
been analyzed using nebular continuum models in conjunction with
modeling of the central star in the UV (\eg,
\citealp{bianchi:97,dopita:97}).  With the \emph{Far-Ultraviolet
Spectroscopic Explorer} (FUSE) , it is now possible to observe LMC
CSPN at far-UV wavelengths where the spectrum is unaffected by nebular
contamination.  Far-UV/UV analyses of Galactic CSPN with winds have
been carried out by \citet{koesterke:98b} and \citet{herald:04b}.  The
far-UV is where these hot stars emit most of their observable flux and
often exhibit their strongest photospheric and/or wind features.
\citet{herald:04c} performed an analysis of the far-UV spectra of the
Galactic CSPN K1-26, and derived a significantly higher temperature
(120 vs. 65~kK) than had resulted from analysis of optical spectra
\citep{mendez:85}, illustrating the value of far-UV-based analyses.

Motivated by the above considerations, we observed a sample of seven
CSPN with FUSE as part of Bianchi's cycle 2 program B001.
The FUSE spectra allow us to separate the nebular and central star components
and to characterize the physical parameters of each through modeling.
The FUSE range also includes strong absorption due to molecular
hydrogen (\Htwo), which is also modeled concurrently.

This paper is arranged as follows.  The observations and data
reduction are described in \S~\ref{sec:obs}.  A comparison of the
spectra of the objects is presented in \S~\ref{sec:description}.  Our
models and parameter determinations are described in
\S~\ref{sec:modeling}.  The implications of our results
are discussed in \S~\ref{sec:discussion} and our conclusions in
\S~\ref{sec:conclusions}.

\section{OBSERVATIONS AND REDUCTION}\label{sec:obs}

For our analysis, we used far-UV data taken with the \emph{Far
Ultraviolet Spectroscopic Explorer} (FUSE) and archive UV data
gathered with the \emph{Hubble Space Telescope's} (HST) 
\emph{Faint Object Spectrograph} (FOS) or \emph{Space Telescope
Imaging Spectrograph} (STIS).

\subsection{Far-UV Data}

The FUSE observations of our sample stars are summarized in
Table~\ref{tab:fuseobs}.  They represent some of the dimmest stellar
objects yet observed by FUSE, and necessitated long integration
times.  

FUSE covers the wavelength range 905---1187~\AA\ at a spectral
resolution of $R \approx 20,000$ ($\sim$ 15 \kms). It is described by
\citet{moos:00} and its on-orbit performance is discussed by
\citet{sahnow:00}. FUSE collects light concurrently in four different
channels (LiF1, LiF2, SiC1, and SiC2), each of which is divided into
two segments (A \& B) recorded by two detectors, covering
different subsets of the above range with some overlap.
We used FUSE's LWRS (30\arcsec$\times$30\arcsec) aperture.  These
data, taken in ``time-tag'' mode, have been calibrated using the most
recent FUSE data reduction pipeline, efficiency curves and wavelength
solutions (CALFUSE v2.2.2).

Because of their faintness, most of our targets required many orbit-long
exposures, each of which typically had low count-rates and thus
signal-to-noise ratios.  Each calibrated extracted sequence was
checked for unacceptable count-rate variations (a sign of detector
drift), incorrect extraction windows, and other anomalies.  
Segments with problems were not included in further steps.  The 2-D
spectra were checked to ensure no secondary objects contaminated the
exposure.

\begin{figure}
\begin{center}
\epsscale{.25}
\rotatebox{270}{
\plotone{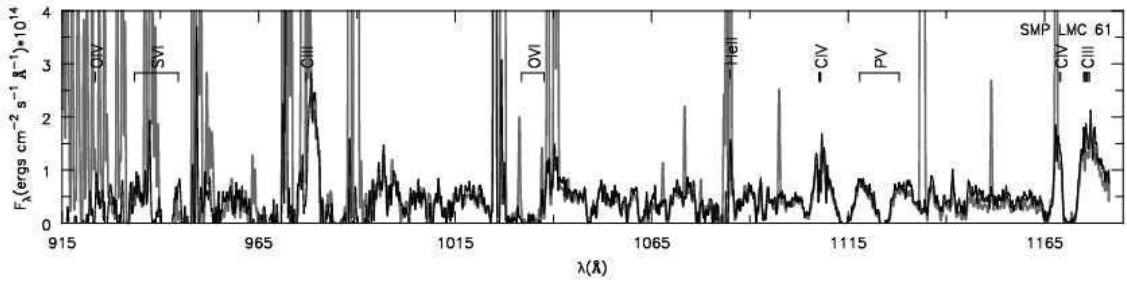}}
\caption{SMP LMC 61 observations: Comparison of the ``night only''
data (black line) with the combined day and night (``all'') data
(gray line).  Strong terrestrial airglow lines (\eg\ \OI) are
seen in many regions of interest in the pipeline extraction of
combined data.  Due to the orientation of FUSE, scattered solar lines
(\eg\ \OVI) appear in the SiC detectors.  The data are convolved with
a 0.25~\AA\ Gaussian for clarity.  }
\label{fig:daynit}
\end{center}
\end{figure}

An example of FUSE observations is shown in Fig.~\ref{fig:daynit}.
Two data reductions are shown for one object: all the ``good'' data
(\ie, not affected by bursts or other anomalies), and only the good
data taken during the night portion of the orbits (when the spacecraft
was on the dark side of the Earth).  The spectrum is contaminated by
both strong terrestrial airglow lines (such as \OI\ and \NI), and
scattered solar features (\OVI\ in this case, and \CIII in others).
The latter appear in the SiC detectors due to the orientation of FUSE.
Because the terrestrial/solar features often confuse the CSPN
observations, we only used the night-only data in our spectral
analysis, and all FUSE spectra henceforth presented in this paper will
be night-only.  Note that some residual airglow lines (usually \OI) still
sometimes appear in this night only data.

All the ``good'' calibrated exposures were combined using FUSE
pipeline routines.  The default FUSE pipeline attempts to model the
background measured off-target and subtracts it from the target
spectrum.  We found that, for our fainter objects, the background
appeared to be over-estimated with this method, particularly at
shorter wavelengths (\ie, $<1000$~\AA).  We therefore tried two other
reductions.  In the first, subtraction of the measured background is
turned off and the background is taken to be the model
scattered-light scaled by the exposure time.  In the second, the first
few steps of the pipeline are run on the individual exposures (which
correct for effects unique to each exposure such as Doppler shift,
grating motions, etc).  The photon lists for the individual exposures
are then combined and the remaining steps of the pipeline run on the
combined file, with the motivation being that more total counts for both
the target and background allow for a better extraction.  However,
this method did not result in a significant improvement over the others.
The adopted background model for each star is indicated in
Table~\ref{tab:fuseobs}.

The resulting calibrated data of the different channels were then
compared for consistency.  There are several regions of overlap
between the different channels, and these were used to ensure that the
continuum matched. For the fainter targets, often the continuum levels
for different detectors did not agree.  The procedure we typically
adopted was to trust the LiF1a absolute flux calibration (which is
most reliable --- \citealp{sahnow:00}), and scale the flux of the
other detectors if needed.  The LiF1b segment is severely affected by
an artifact known as ``the worm'' (FUSE Data Handbook v1.3) and was
not used, except in one case where it was in good agreement with the
LiF2a.  The SiC1 detector seemed frequently off-target, and other
segments offered higher S/N data in the same range, so its data was
often not used.  In the end, we typically used the SiC2 data for
wavelengths below $\sim1000$~\AA.  Longwards of 1080~\AA\ LiF2a data
was employed.  For the intermediate region (\ie, 1000--1080~\AA), data
from LiF1b, LiF2b, or both were used (data from one segment was
omitted if it was discrepant with that of the SiC detectors).  The
region between 1083 and 1087~\AA\ is not covered by the LiF detectors,
and as the SiC detectors in this range were off-target, we have
omitted this region (appearing as gaps in the observed data).
Table~\ref{tab:fuseobs} shows which segments and portions were used
for the combined spectrum.

The results from the three different reduction methods were compared
during the above process to understand if some features were actually
artifacts of the background subtraction process.  Data from near the
detector edges were also omitted if they looked inconsistent.

The FUSE spectra presented in this paper were obtained by combining the
good data from different segments, weighted by detector sensitivity, and
rebinning to a uniform dispersion of 0.05~\AA.

\subsection{UV Data}

Archival data from the HST's \emph{Faint Object Spectrograph} (FOS)
and \emph{Space Telescope Imaging Spectrograph} (STIS) instruments
were also used longwards of 1200~\AA, as summarized in
Table~\ref{tab:uvobs}.  The FOS archive data were taken in the
1\arcsec\ aperture, whose actual diameter following the COSTAR
installation is 0.86\arcsec.  The angular sizes of the LMC PN are
typically $\lesssim$1\arcsec, so the FOS spectroscopy typically
includes the entire nebula (the exception is SMP LMC~35 see
Table~\ref{tab:neb}).  The utilized FOS dispersers include the G130H
($\res\sim1$~\AA), the G190H ($\res\sim1.5$~\AA), the G270H
($\res\sim2$\~\AA) and the G400H ($\res\sim3$~\AA).  For SMP LMC~61,
high-resolution STIS data was available, taken through the 52\arcsec
x0.2\arcsec\ aperture with the G140L ($\res\sim1.3$~\AA), G230L
($\res\sim1.5$~\AA) and G430L ($\res\sim1.6$~\AA) gratings.

Generally, the flux levels of the FUSE and FOS/STIS data in the region
of overlap are in good agreement.  However, for SMP LMC~2, the FOS
flux levels were about 60~\% those of FUSE.  There was nothing in the
FOS log files that suggested the observation may have been slightly
off-target.  We looked at the FUSE visitation images of the field
during, and there is a UV-bright source that could have been within
the FUSE aperture during the observation.  Since FUSE has no spatial
resolution, this source could have contaminated the observation.  We
assume this is the source of the discrepancy, and thus use the FOS
flux level when scaling our parameters in the later analysis.

\section{Description of Spectra}\label{sec:description}

\begin{figure}
\begin{center}
\epsscale{1}
%\begin{turn}{90}
\rotatebox{0}{
\plotone{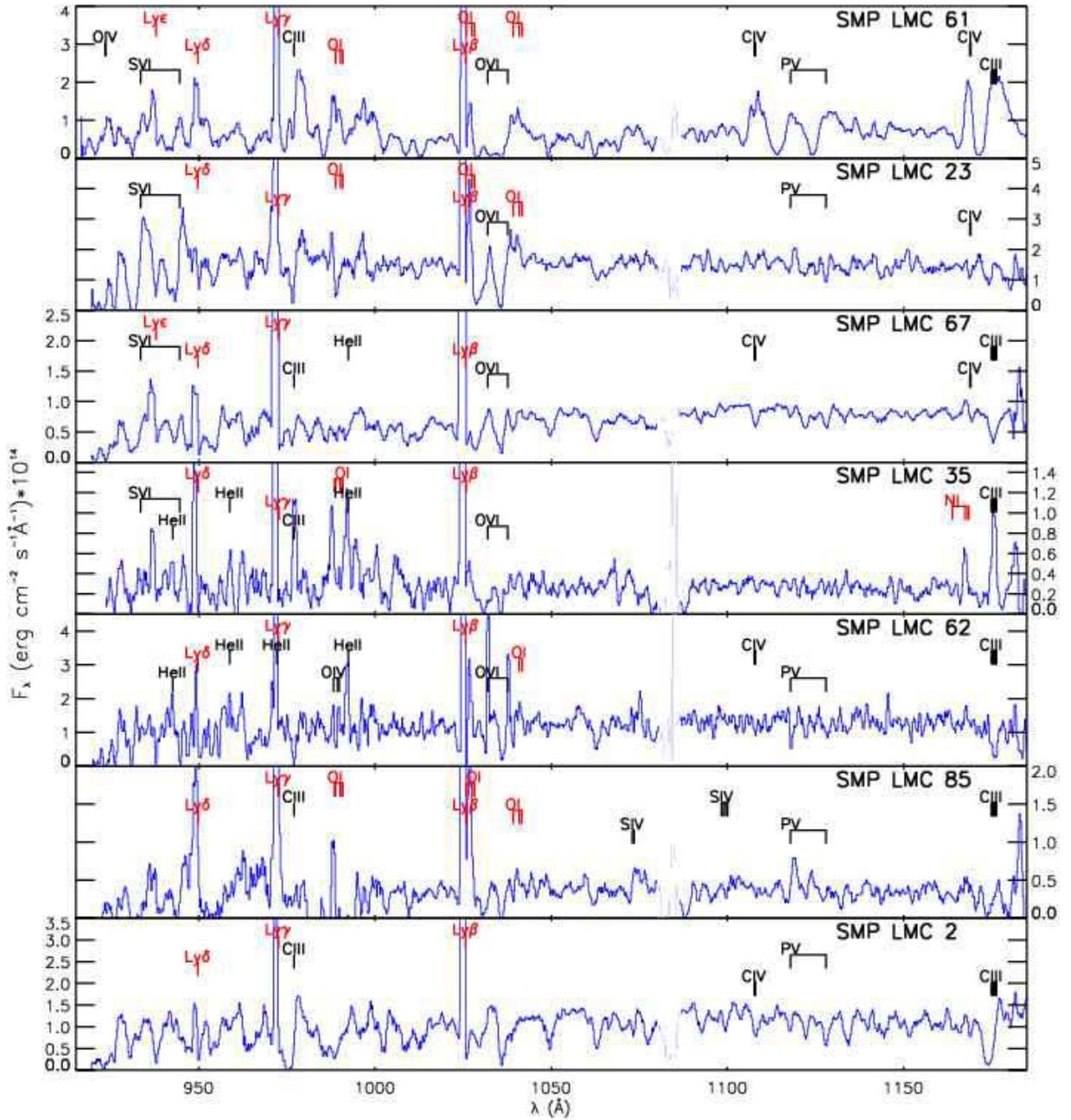}}
%\end{turn}
\caption{FUSE spectra of our sample, velocity shifted to the rest frame of
  the central star (Table~\ref{tab:photo}), and rebinned to a
  resolution of 0.25\AA.  The more prominent stellar features and nebular
  emission lines are marked by black labels, airglow features are marked by
  the red/gray labels.  Most objects display wind lines to some extent.
  \CIII\ features (either in absorption or P-Cygni) appear in most.
  The \OVI\ doublet seems to be present in all cases except for 
  SMP LMC~2, 62, and 85, where it is not obvious.  Note the \OVI\ \doublet
  1032-38 nebular emission features of SMP LMC~62, probably originating from
  shocks.}\label{fig:lmc_fuse}
\end{center}
\end{figure}

\begin{figure}
\begin{center}
\epsscale{1}
%\begin{turn}{90}
\rotatebox{0}{
\plotone{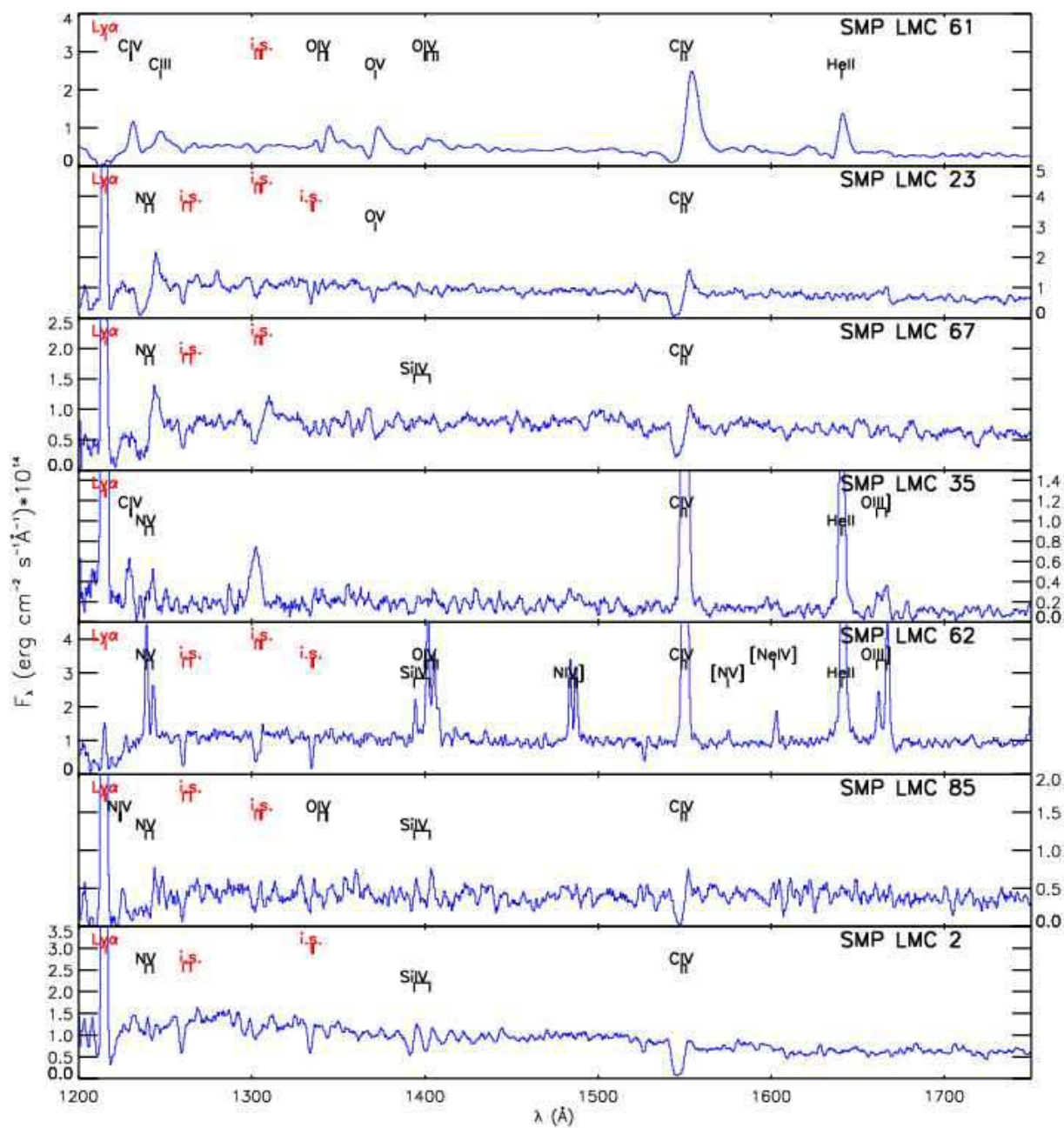}}
%\end{turn}
\caption{UV spectra (FOS and STIS) of our sample, velocity shifted to
  the rest frame of the central star (Table~\ref{tab:photo}).
  The more prominent stellar features and
  nebular emission lines are marked with black labels, airglow lines or
  interstellar absorption features are marked with red/gray labels.  Most
  objects display wind lines to some extent.  \CIV\ \doublet 1548-51
  appears as a P-Cygni profile in most.  SMP LMC~62 shows a rich emission
  line spectrum typical of a shocked region.}\label{fig:lmc_uv}
\end{center}
\end{figure}

The reduced FUSE spectra for the entire sample, along with line
identifications, are shown in Fig.~\ref{fig:lmc_fuse}, and the UV
spectra are shown in Fig.\ref{fig:lmc_uv}.  In Table~\ref{tab:lines},
we list lines common in our sample, and classify them based on whether
they appear as P-Cygni profiles, absorption lines, or nebular emission
features.  If a line appears as a strong P-Cygni profile and the blue
edge of the absorption trough was not obviously obscured by other
features (airglow lines or strong interstellar absorption features),
we measured the terminal velocity based on its blue edge.  This
measurement, \vedge, gives a fairly good indication of the wind
terminal velocity, \vinf, the difference between the two being the
turbulent Doppler broadening, which is typically $\sim$10-20~\%.

For most of the sample stars  with wind signatures, $700
\lesssim \vedge \lesssim 1300$~\kms.  In the far-UV especially, \Htwo\
absorption makes the continuum level very difficult to set
(\S~\ref{sec:htwo}).  Thus some measurements are more reliable than
others.  We used these measurements to initially estimate \vinf\
for our models (\S~\ref{sec:diag}).

Five out of seven objects of our sample display definite wind
signatures.  SMP LMC~62 has questionable \OVI\ wind features, and its
UV lines are contaminated by nebular emission features.  The FUSE
spectra of SMP LMC~35, the dimmest object, has the lowest S/N and is
not of sufficient quality to make a definitive statement about the
existence of a wind.  In its UV spectrum, nebular emission features
again hide any present stellar features.  There is a feature at $\sim
1300$~\AA\ that could be a \CIII\ P-Cygni line, the strong \OI\
interstellar absorption feature at 1300~\AA\ is confusing the issue.
We now discuss the individual spectra.

The UV spectra of SMP LMC~23 and
LMC~67 are almost identical, both showing 
prominent wind lines of \NV\ \doublet 1238-43 and \CIV\ \doublet
1548-51.  The blue edge of the \CIV\ doublet indicates
similar terminal velocities ($\simeq 1100$~\kms)
The far-UV spectra are likewise similar, with the wind lines
(\SVI\ \doublet 933-44, \CIII\ \singlet 977, \OVI\ \doublet 1032-38)
of SMP LMC~23 being slightly stronger.  A noticeable difference is \CIII\
\singlet 1175, which is in absorption in SMP LMC~67 and filled in in
SMP LMC~23.

\begin{figure}
\begin{center}
\epsscale{.25}
\rotatebox{270}{
\plotone{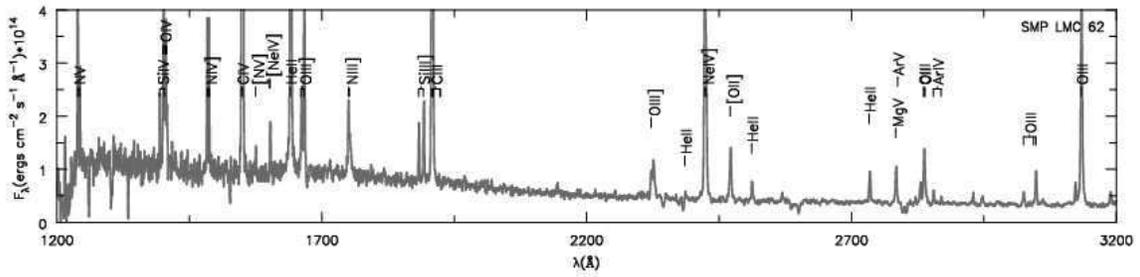}}
\caption{SMP LMC~62: The far-UV/UV spectrum of
  this object displays many highly-ionized emission lines
  characteristic of a shocked environment, such as \NV\ \doublet 1238-43
  and [\NeV ] \singlet 1146.
}
\label{fig:lmc62_uv}
\end{center}
\end{figure}

Fig.~\ref{fig:lmc62_uv} shows its far-UV and near-UV spectrum of
SMP LMC~62.  This star is unique among our sample, showing \OVI\ \doublet
1032-38 nebular features in its far-UV spectrum, as well as many
high-ionization emission lines in its UV spectrum (such as [\NeV]).
Table~\ref{tab:lmc62uv} lists the measured emission line fluxes for
this object, determined by trapezoidal integration with suitable
continuum subtraction.

The far-UV spectra of SMP LMC~62 is similar to that of SMP LMC~67, both
showing weak (if any) \SVI\ \doublet 933-44, \CIII\ \singlet 977 and
\OVI\ \doublet 1032-38.  \CIII\ \singlet 1175 appears in absorption in
both.  One key difference is that SMP LMC~62 displays what appear to
be nebular \OVI\ emission features, which are unique in our sample.
Such features have also been observed in the FUSE spectrum of the
Galactic CSPN NGC~2371 \citep{herald:04b}.
The UV emission features mask \CIV\ \doublet 1548-51 if present.  \NV\
\doublet 1238-43 may be a wind profile, but the nebular emission again
confuses the issue.

SMP LMC~61 is a [WC] star, displaying a rich spectrum of Carbon and Oxygen
P-Cygni features.  There are no obvious nitrogen lines in its spectrum.
SMP LMC~2 and SMP LMC~85 display \CIII\ in their far-UV spectra and have similar
UV spectra, with the \NV, \SiIV, and \CIV\ features of SMP LMC~85 being
more prominent.

\section{MODELING}\label{sec:modeling}

Modeling the central stars of compact planetary nebulae presents some
challenges.  First, their optical spectra (and frequently their UV
spectra also) are contaminated by the nebular continuum and lines,
which obscure and sometimes entirely mask the stellar spectrum in
these regions.  Thus, one must rely on a smaller set of stellar
features in the far-UV and UV to determine the stellar parameters.
Second, the far-UV region, while not affected by nebular continuum
emission, is affected by absorption by electronic transitions of
sight-line molecular hydrogen (\Htwo).  In our objects, such
absorptions affect the entire far-UV region.  Thus the analysis of these
spectra consists of modeling the central star spectrum, the nebular
continuum emission longwards of 1200~\AA, and the sight-line Hydrogen
(atomic and molecular) shortwards concurrently and to find a
consistent solution.  However, we discuss each in turn for clarity.

Throughout this paper, we adopt a LMC distance of $D=50.6$~kpc
\citep{feast:91}, and an LMC metallicity of $z = 0.4\Zsun$
\citep{dufour:84} with values of ``solar'' abundances from
\citet{gray:92}.

\subsection{Molecular and Atomic Hydrogen}\label{sec:htwo}

Absorption due to atomic (\HI) and molecular hydrogen (\Htwo) along the
sight-line complicates the far-UV spectra of these objects.
Toward a CSPN, this sight-line material typically consists of
interstellar and circumstellar components.  Material comprising the
circumstellar \HI\ and \Htwo\ presumably was ejected from the star
earlier in its history, and is thus important from an evolutionary
perspective.

\begin{figure}
\begin{center}
\epsscale{1}
%\begin{turn}{90}
\rotatebox{0}{
\plotone{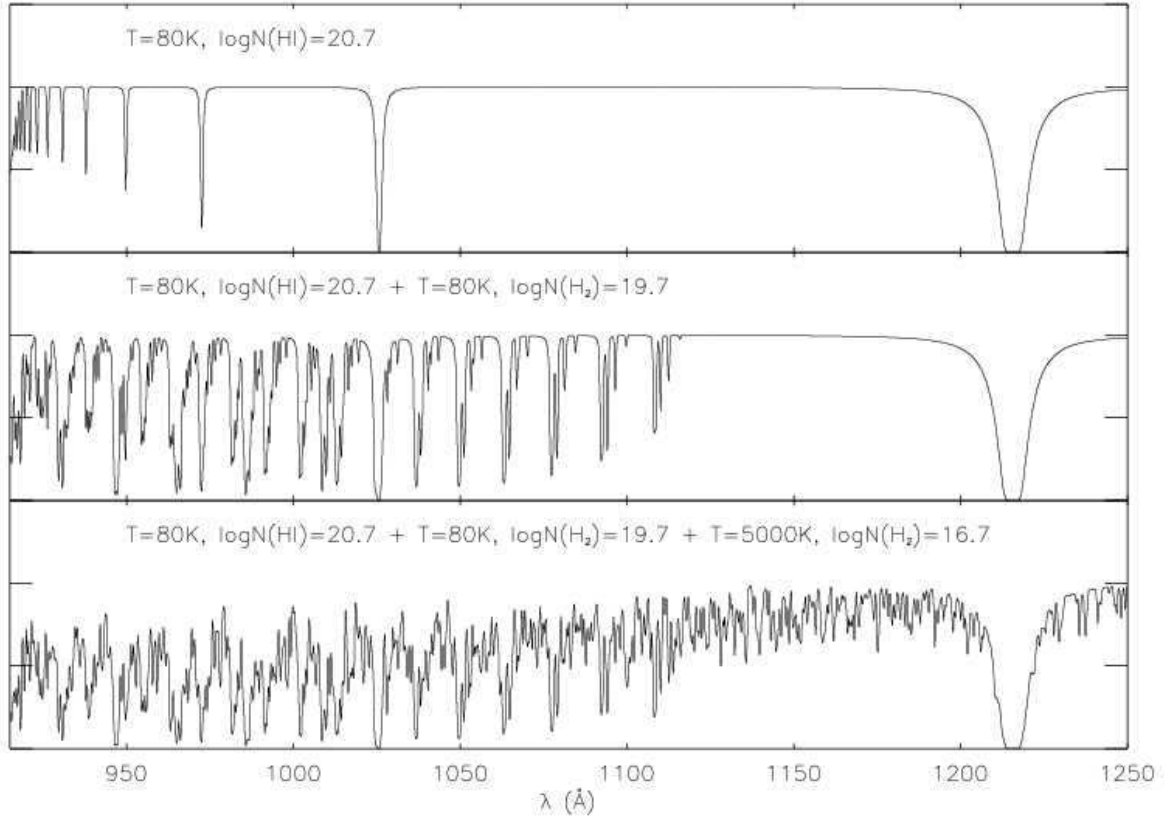}}
%\end{turn}
\caption{\Htwo\ effects: The top plot shows a flat continuum to which
  are applied the absorption effects of an atomic hydrogen gas with
  characteristics of the ISM along a sight-line of low reddening [\ie,
  $T=80$~K, column density of $\logN(\HI) = 20.7$~cm$^{-2}$], typical
  for our targets.  The middle plot shows the absorption pattern of a
  typical low-reddening (\EBMV=0.1) interstellar \Htwo\ gas [$T=80$,
  $\logN(\Htwo) = 19.7$~cm$^{-2}$], added to the previous.
  The bottom plot shows the effects of a relatively small quantity
  [$\logN(\Htwo^{hot})=16.7$~cm$^{-2}$] of hot ($T=5000$~K) \Htwo,
  similar to what we observe toward many of our LMC CSPN, applied to
  the previous.  The entire FUSE range is affected by the dense field
  of transitions of numerous ro-vibrational \Htwo\ states, which suppress the
  far-UV continuum.}
\label{fig:htwo}
\end{center}
\end{figure}

The \Htwo\ transitions in the FUSE range are from the Lyman ($B^1
\Sigma^+_u$--$X^1 \Sigma^+_g$) and Werner ($C^1 \Pi^{\pm}_u$--$X^1
\Sigma^+_g$) series.  For a ``cool'' molecular gas (\ie, $\lesssim
100$~K) as found in the ISM, only the ground vibrational state is
populated and the absorptions associated with its different rotational
states show up as a series of discrete absorption ``bands'', confined
shortwards of $\lesssim 1115$~\AA\ for typical IS column densities
(\ie, $10^{18}$---$10^{19}$cm$^{-2}$) (see Fig.~\ref{fig:htwo}).  The
effects of such a gas are not difficult to correct for, because one
can identify and use the unaffected parts of the stellar spectrum as a
reference.  However, we find that for many objects of our sample, a
significant amount of \emph{hot} (\ie, $T\gtrsim1000$~K) \Htwo\ lies
along their sight-lines, presumably associated with the nebula.  At
such temperatures, higher vibrational states become populated and the
absorption pattern is much more complex (again, see
Fig.~\ref{fig:htwo}), obscuring the entire far-UV continuum spectrum.
With care, the effects of such a gas can be accounted for and stronger
stellar features can be discerned, but weaker stellar features will be
unrecoverable.

\begin{figure}
\begin{center}
\epsscale{1}
%\begin{turn}{90}
\rotatebox{0}{
\plotone{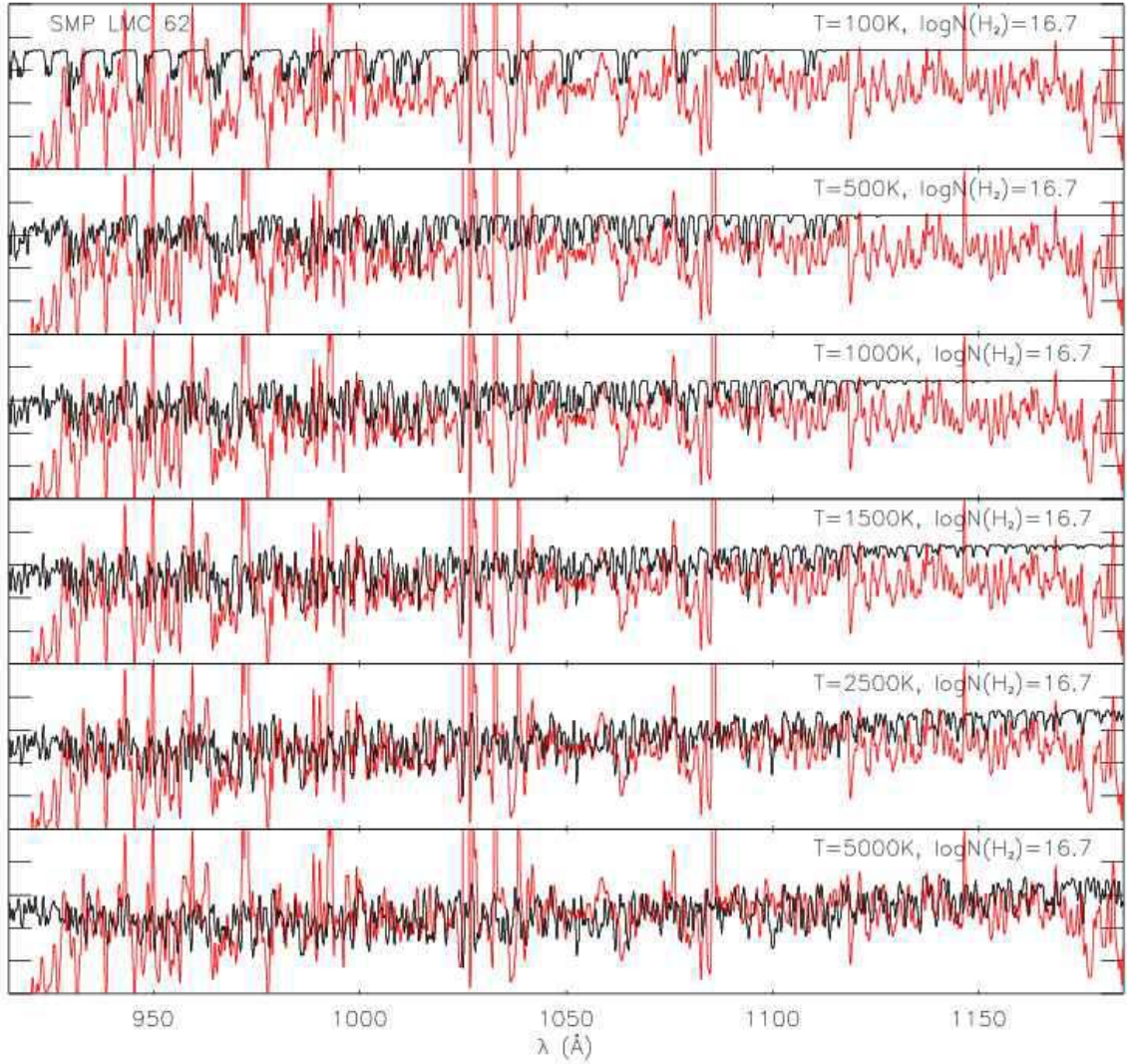}}
%\end{turn}
\caption{SMP LMC 62: Fitting hot \Htwo.  The FUSE observations (red/gray)
  are shown, along with \Htwo\ absorption models of different
  temperatures ($\log{N}=16.7$~cm$^{-2}$) applied to flat continuum
  (black).  The absorption pattern implies a very hot \Htwo\ gas of
  $T\gtrsim 2500$~K.  Such high temperatures are characteristic of a
  shocked environment (discussed in \S~\ref{sec:results})}\label{fig:lmc62_h2}
\end{center}
\end{figure}

We have modeled the \Htwo\ toward our sample CSPN in the following
manner.  For a given column density ($N$) and gas temperature ($T$),
the absorption profile of each line is calculated by multiplying the
line core optical depth ($\tau_0$) by the Voigt profile $[H(a,x)]$
(normalized to unity) where $x$ is the frequency in Doppler units and
$a$ is the ratio of the line damping constant to the Doppler width
(the ``b'' parameter).  The observed flux is then $F_{obs} =
\exp{[-\tau_{0}H(a,x)]} \times F_{intrinsic}$.  We first assume the
presence of an interstellar component with $T=80$~K (corresponding to
the mean temperature of the ISM --- \citealp{hughes:71}) and \vturb =
10\kms.  The column density of this interstellar component is
estimated by fitting its strongest transitions.  If additional
absorption features due to higher-energy \Htwo\ transitions are
observed, a second (hotter) component is modeled, presumed to be
associated with the nebula.  We thus velocity-shift the \Htwo\
absorption features of the circumstellar component to the radial
velocity (from Table~\ref{tab:mod_param}) of the particular star.  The
temperature of the circumstellar component can be determined by the
absence/presence of absorption features from transitions of more
energetic vibrational and rotational states, and the column density by
fitting these features (an example of the sensitivity of the \Htwo\
spectrum to the temperature is shown in Fig.~\ref{fig:lmc62_h2}).
Iterations between fitting the interstellar and circumstellar
components may be required, as both contribute to the lower-energy
features, but it is not possible to separate the components.  We note
that our terminology of ``circumstellar'' and ``interstellar''
components is a simplification, and basically indicate that the
``cool'' component is assumed mostly interstellar and the ``hot''
component is assumed mostly circumstellar.  However, the column
density derived for the cooler ``interstellar'' component may also
include circumstellar \Htwo.

When possible, \HI\ column densities are determined from the \Lya\ and
\Lyb\ features, which encompass both the interstellar and nebular
components.  The absorption profiles of \HI\ are calculated in
a similar fashion as described above for \Htwo.

Our derived parameters for sight-line atomic and molecular Hydrogen
are shown in Table~\ref{tab:hydrogen}.  All objects require very high
temperatures to fit the \Htwo\ absorption patterns, ($T \gtrsim
2000$~K).  The absorption pattern become less sensitive to temperature
at high values, but is sensitive to the adopted column density.
Molecular hydrogen disassociates around $T \simeq 2500$, so such high
temperatures are signs of non-equilibrium conditions (note our models
assume thermal populations).  Such \Htwo\ characteristics are observed
in shocked regions.  We will discuss this topic more in
\S~\ref{sec:discussion}.  A shocked region is likely complicated,
containing areas of \Htwo\ of differing characteristics.  Because of
the high complexity of the absorption pattern at these hot
temperatures, we did not attempt to fit every feature, but the
absorption pattern in the far-UV as a whole.

We also list in Table~\ref{tab:hydrogen} the reddenings implied by our
measured column densities of \HI\ using the relationship $\left<
N(\HI)/\EBMV \right> = 4.8\E{21}$~cm$^{-2}$~mag$^{-1}$
\citep{bohlin:78}, which represents typical conditions in the ISM.
These column densities indicate higher reddenings than those derived
from the logarithmic extinction (Table~\ref{tab:neb}) or from our
model fits (Table~\ref{tab:mod_param}).  This is likely explained by
the measured column density including a significant amount of
\emph{circumstellar} \HI, which apparently has a smaller dust-to-\HI\
ratio than that of the ISM.

Thanks to FUSE observations of CSPN, it is becoming apparent that it
is not uncommon to find hot circumstellar \Htwo\ around CSPN.  \Htwo\
with $T < 2000$~K has been observed in the far-UV spectra of Galactic
CSPN by \citet{herald:02} and \citet{mccandliss:00}.  The extreme
temperatures seen in our sample is probably related to their early
evolutionary stage in post-AGB evolution.

\subsection{The Nebular Continuum}\label{sec:nebcont}

Nebular parameters for the program objects taken from the literature
are compiled in Table~\ref{tab:neb}.  They include the angular size of
the nebula $\theta$, the nebular radii $r_{neb}$, the expansion
velocity $v_{exp}$, the electron density $n_{e}$, the electron
temperature $T_e$, the \Hbeta\ flux $F_{H\beta}$, the Helium to
Hydrogen ratio and the doubly to singly ionized Helium ratio.  Values
in \textbf{bold} are our derived values (see below).  We also list the
reddening values determined from literature measurements of the
logarithmic extinction (from H$\beta$) using the relation $c_{H\beta}
= 1.475 \EBMV$.

The nebular continuum significantly contributes to the flux at
wavelengths $\gtrsim 1200$~\AA\ and must be estimated to fit the
stellar spectra (\S~\ref{sec:cs}).  The reddening of our sample CSPN
is typically small, and so it is not a significant source of
uncertainty.

\begin{figure}
\begin{center}
\epsscale{1}
%\begin{turn}{90}
\rotatebox{0}{
\plotone{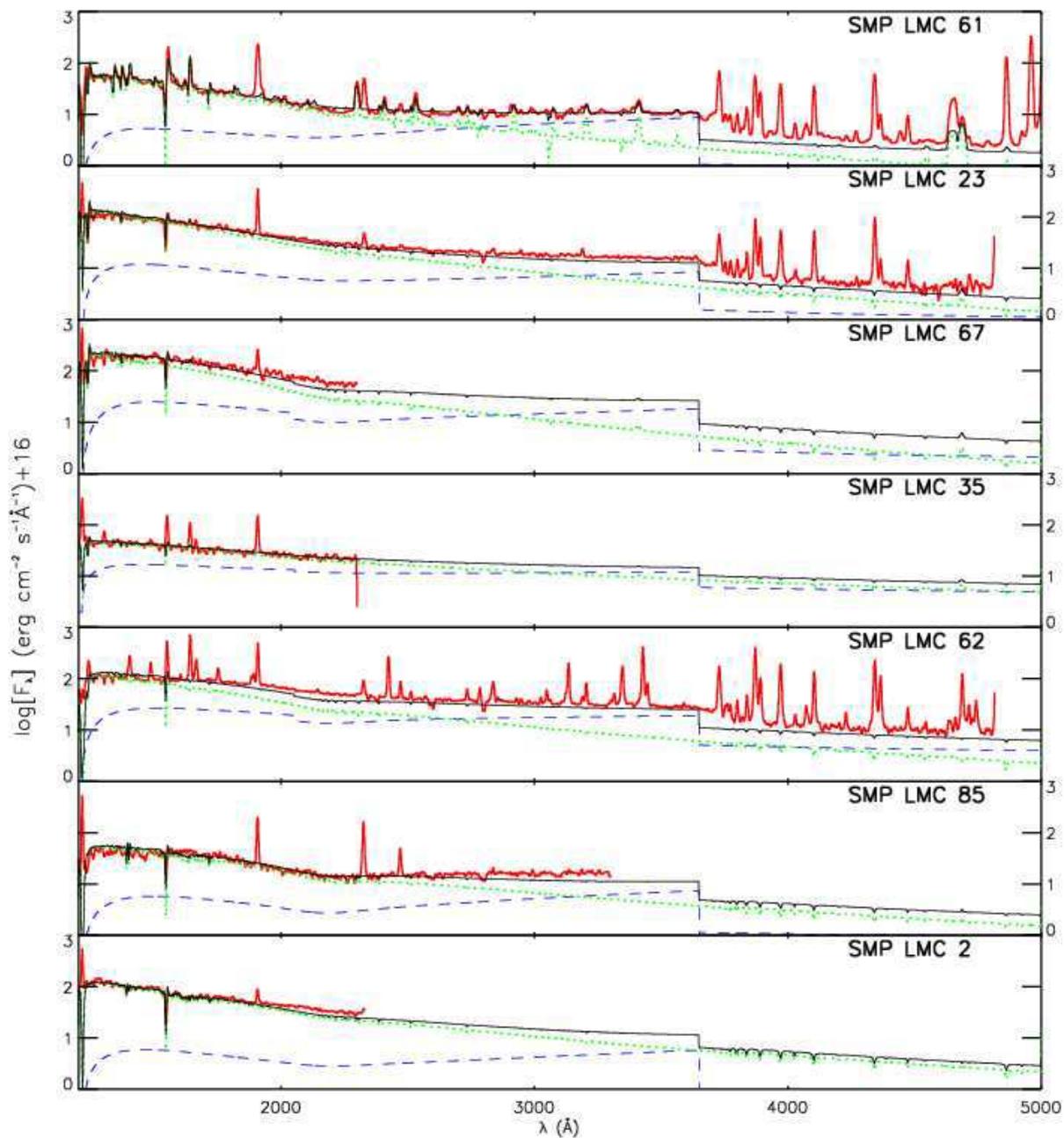}}
%\end{turn}
\caption{The observations (red/gray) are shown along with our stellar
  (green/dotted) and nebular continuum models (blue/dashed).  The sum
  of the models is shown by the black solid line.  All models have been
  reddened with our determined values for \EBMV\
  (Table~\ref{tab:mod_param}), and the effects of hydrogen absorption
  (Table~\ref{tab:hydrogen}) have been applied.  The spectra have been
  convolved with a 5~\AA\ Gaussian for clarity.  Note the logarithmic
  flux scale.  }\label{fig:neb}
\end{center}
\end{figure}

The nebular continuum emission has been modeled using the code
described in \citet{bianchi:97}, which accounts for two-photon, H and
He recombination, and free-free emission processes.  The computed
emissivity coefficient of the nebular gas is scaled as an initial
approximation to the total flux at the Earth by deriving the emitting
volume from the dereddened absolute \Hbeta\ flux. In the cases where
the spectroscopic aperture was smaller than the nebular size (SMP
LMC~35 and 61) the nebular flux was scaled by the corresponding
geometrical factor.  $T_e$ and \nelec\ from the literature are used as
initial inputs, and adjusted if necessary.  SMP LMC~2 and 85 have no
published values of the electron density.  For these, we tried
different values as inputs and found $\nelec = 5000$ and 40,000
cm$^{-3}$, respectively, to produce satisfactory fits.  The high
\nelec\ for SMP LMC~85 is consistent with its very compact morphology
($r_{neb} \le 0.2$~pc).  The values of $F_{H\beta}$ and c$_{H\beta}$
are perhaps the most uncertain, and finally we re-scaled the nebular
flux with the assumption that the nebular continuum is responsible for
the majority of flux at longer wavelengths (the $H\beta$ fluxes
implied by our scaling are listed in \textbf{bold} in the table).
This assumption produced an overall good fit of the observed spectra
over the entire wavelength range.  The nebular continuum is
constrained not to exceed the P-Cygni troughs in the UV range.  Values
of $F_{H\beta}$ listed in parentheses reflect this scaling.  Our
combined stellar and nebular models, along with the observations, are
shown in Fig.~\ref{fig:neb}.

\subsection{The Central Stars}\label{sec:cs}

As discussed in previous sections, the modeling of the central stars
is done concurrently with that of the nebular continuum
(\S~\ref{sec:nebcont}) and the hydrogen absorption
(\S~\ref{sec:htwo}).  The hot molecular hydrogen along the sight-lines
of many of these objects makes the identification of the stellar
features in the far-UV and their analysis challenging.  For
wavelengths shorter than 1200~\AA, the nebula does not contribute
continuum emission, and this region can be used to set the stellar
radius, since the distance is well-established.  The reddening has to
be derived concurrently, but it is always small, and therefore it does
not contribute much to the uncertainties.

\subsubsection{The Models}

Due to the severe \Htwo\ absorption in the far-UV, there are few
photospheric absorption lines are available, so we have relied primarily
on wind features to determine the parameters of the CSPN.
Intense radiation fields, a (relatively) low wind density, and an
extended atmosphere invalidate the assumptions of thermodynamic
equilibrium and a plane-parallel geometry for our sample.  To model
the winds of these objects, we use the non-LTE (NLTE) line-blanketed
code CMFGEN of \citet{hillier:98,hillier:99b}.  CMFGEN solves the
radiative transfer equation in an extended, spherically-symmetric
expanding atmosphere.  Originally developed to model the winds of
(massive) Wolf-Rayet stars, it has been adapted for objects with
weaker winds such as O-stars and CSPN as described in
\citet{hillier:03}.

The detailed workings of the code are explained in the references
above.  In summary, the code solves for the NLTE populations in the
comoving-frame of reference.  The fundamental photospheric/wind
parameters include \Teff, \Rstar, \Mdot, the elemental abundances and
the velocity law (including the wind terminal velocity, \vinf).  The
\emph{stellar radius} (\Rstar) is taken to be the inner boundary of
the model atmosphere (corresponding to a Rosseland optical depth of
$\sim20$).  The temperature at different depths is determined by the
\emph{stellar temperature} \Tstar, related to the luminosity and
radius by $L = 4\pi\Rstar^2\sigma\Tstar^4$, whereas the
\emph{effective temperature} (\Teff) is similarly defined but at a
radius corresponding to a Rosseland optical depth of 2/3 (\Reff).  The
luminosity is conserved at all depths, so $L =
4\pi\Reff^2\sigma\Teff^4$.  

We assumed what is essentially a standard velocity law $v(r) =
\vinf(1-r_0/r)^\beta$ where $r_0$ is roughly equal to \Rstar.  The
choice of the velocity law mainly affects the profile shape, not the total
optical depth, and does not greatly influence the derived
parameters.  Once a velocity law is specified, the density structure
of the wind $\rho(r)$ is then parameterized by the mass-loss rate
\Mdot\ through the equation of continuity: $\Mdot=4\pi\Rstar^2
\rho(r)v(r)$.

It has been found that models with the same \emph{transformed radius}
\Rtrans\ [$\propto \Rstar(\vinf/\Mdot)^{2/3}$] \citep{schmutz:89} and
\vinf\ have the same ionization structure, temperature stratification
(aside from a scaling by \Rstar) and spectra (aside from a scaling of
the absolute flux by $\Rstar^2$ --- \citealp{schmutz:89,hamann:93}).
Thus, once the velocity law and abundances are set, one parameter may
be fixed (say \Rstar) and parameter space can then be explored by
varying only the other two parameters (\eg, \Mdot\ and \Teff).
\Rtrans\ can be thought of as an optical depth parameter, as the
optical depth of the wind scales as $\propto \Rtrans^{-3}$, for
opacities which are proportional to the square of the density.
Scaling the model to the observed flux yields \Rstar/$D$.  An LMC
distance of $D = 50.6$~kpc was adopted to determine \Rstar.

\subsubsection{Clumping}\label{sec:clumping}
Radiation driven winds have been shown to be inherently unstable
\citep{owocki:88,owocki:94}, which should lead to the formation of
clumps.  The degree of clumpiness is parametrized by $f$, the
\emph{filling factor}.  One actually can only derive
\Mdots=(\Mdotc/$\sqrt{f}$) from the models, where \Mdots\ and \Mdotc\
are the smooth and clumped mass-loss rates.  For the denser winds of
population I Wolf-Rayet stars, the clumping factor can be constrained
by the strength of the electron scattered line wings, and clumping
factors of $f \sim 0.1$ are typical (a reduction of \Mdot\ to roughly
a third of its unclumped value).  For O-stars, the lower mass loss
rates make the electron scattering effects small and difficult to
ascertain \citep{hillier:03}.  The winds of our sample stars are even
weaker.  Given that the wind lines in the far-UV are affected by \Htwo,
and the UV FOS data are not high resolution, we did not attempt to
rigorously constrain the degree of clumping in the winds of our
sample.  We calculated test models with different clumping factors,
and generally found the spectrum to be insensitive, except for \OVI\
\doublet\ 1032-38 in the hotter models ($\Teff \gtrsim 50$~kK) and
\PV\ \doublet 1118-28 for cooler models.  In some cases, better
results were achieved, and such cases are discussed in
\S~\ref{sec:results}.  Unless otherwise noted, we have adopted $f=1$
and use \Mdot\ to refer to the smooth mass loss rate throughout this
paper.  This is an upper limit, and the lower limit of \Mdot\ is
estimated to be a third of this value.

\subsubsection{Gravity}\label{sec:gravity}

Because of the severe \Htwo\ absorption in the far-UV, and the masking of
the stellar flux by the nebular continuum at longer wavelengths, there
are no suitable absorption lines to be used as gravity diagnostics.
In CMFGEN, gravity enters through the scale height $h$ ($\propto
g^{-1}$), which connects the spherically extended hydrostatic outer
layers to the wind.  The relation between $h$ and $g$, defined in
\citet{hillier:03}, involves the mean ionic mass and mean number of
electrons per ion, the local electron temperature and the ratio of
radiation pressure to the gravity.  Our models typically have scale
heights which are equivalent to $\logg \sim 4.7-5.0$.  Their
hydrostatic structures were verified with test models generated using
\tlusty\ \citep{hubeny:95}, which calculates the atmospheric structure
assuming radiative and hydrostatic equilibrium, and a plane-parallel
geometry, in NLTE conditions. For objects evolving along the
constant-luminosity branch and with $40 \lesssim \Teff \lesssim 70$~kK
(the regime of our objects), evolutionary codes typically adopt $4.0
\le \logg \le 5.3$ (\eg, \citealp{marigo:02}).  Over such gravities,
the derived wind parameters are not that sensitive to $h$.  Spectral
wind features can be fit with the models of the same \Teff\ but with
gravities differing by over a magnitude.  In such cases, the lower
gravity model requires a higher luminosity, as its more extended
atmosphere results in larger discrepancy between \Tstar\ and \Teff.
To illustrate the impact of adopting a different \logg\ on the derived
parameters, a model with $\Teff = 55$~kK, $\logg \simeq 5$, $L \simeq
5000$~\Lsun, and $\Rstar \simeq 0.8$~\Rsun\ produces about the same
spectrum as a model with the same \Teff, but with $\logg=4$, and $L$
about 8~\% larger and \Rstar\ about 17~\% smaller.  For gravities
stronger than $\logg \simeq 5.0$, the discrepancy is smaller, for
lower gravities, larger.  For $\logg=3.7$, \Rstar\ is $\sim30$~\%
smaller and and $L$ is about $\sim20$~\% greater.  Thus, for our
hotter objects (\ie, $\gtrsim 50$~kK) we estimate the errors due to
uncertainty of the gravity to be $\lesssim 20$~\% for the radius and
$\lesssim 10$~\% for the luminosity.  For cooler objects, the errors
are estimated to be $\lesssim 30$~\% for the radius and $\lesssim
20$~\% for the luminosity.

\subsubsection{Abundances}\label{sec:abundances}

CSPN can be generally divided into two groups based on surface
abundances: Hydrogen-rich (those that show obvious Hydrogen lines in
their spectra, and exhibit ``normal'' abundances) and
Hydrogen-deficient (which don't).  The H-rich objects are thought to
be in a quiescent Hydrogen-burning phase, and the H-deficient stars in
a post-Helium flash, Helium-burning phase.  About 10-20~\% of CSPN are
H-deficient (\citealp{demarco:02,koesterke:98b} and references
therein).  This class includes [WC] objects.  It seems the two groups
represent different ``channels'' of CSPN evolution, terminating with
either H-rich or He-rich white dwarfs (DA or DO) (see
\citealp{napiwotzki:99} for a discussion).

To model these CSPN, we constructed two grids of models, ranging in
\Teff\ from 30--80~kK and in $\Rtrans \sim$ 3--200~\Rsun.  The first,
appropriate for ``H-rich'' CSPN, have solar abundances for H and He
but assume an LMC metallicity of $z = 0.4$~\Zsun\ for the metals.  For
the second, corresponding to ``H-deficient'' (or H-poor) CSPN, we have
considered the following.  It has been shown that, throughout the [WC]
subclasses, the spectroscopic differences are mainly tied to \Mdot\
and \Teff, rather than to the abundances \citep{crowther:98,crowther:02}.
The abundances throughout the [WC] subclasses are about constant, with
typical values of (by mass) \XHe = 0.33--0.80, \XC =
0.15--0.50, \XO = 0.06--0.17 \citep{demarco:01}.  The abundance
patterns of PG~1159-[WC] and PG~1159 stars, the likely descendants of
the [WC] class, are similar \citep{werner:01}.  For these
reasons, we have assumed for the second grid a similar abundance
pattern of \XHe/\XC/\XO = 0.55/0.37/0.08 and an LMC
metallicity of $z = 0.4$~\Zsun\ for the other elements (\eg, S, Si, \& Fe).
The nitrogen abundance in these objects typically ranges from none to
$\sim$2\% in Galactic objects, we have adopted $\sim$1\% (by mass).
The abundances for the H-rich and H-deficient grids summarized in
Table~\ref{tab:abund}.  We note here that throughout this work, the
nomenclature $X_i$ represents the mass fraction of element $i$, and
`\Xsun'' denotes the solar value.

Strictly speaking, ``H-rich'' CSPN are those which display detectable
hydrogen in their spectra.  Even in the optical, hydrogen abundances
are often difficult to measure in these objects because hydrogen
features are often masked by those of \HeII\ (see
\citealp{leuenhagen:98} for a discussion).  We emphasize here that
there are no diagnostics in the far-UV or UV spectra of our objects that
allow us to make a statement regarding their hydrogen abundance.
The handful of wind lines upon which we are basing our parameter
determinations do not allow us to make firm statements regarding their
abundances.  

For the model ions, CMFGEN utilizes the concept of ``superlevels'',
whereby levels of similar energies are grouped together and treated as
a single level in the rate equations \citep{hillier:98}.
Ions and the number of levels and superlevels included in the model
calculations, as well references to the atomic data, are given in the
Appendix (\S~\ref{sec:atomic}).

\subsubsection{Diagnostics}\label{sec:diag}

The terminal velocity (\vinf) can be estimated from the blue edge of
the P-Cygni absorption features (Table~\ref{tab:lines}), preferably
those that form further out in the wind.  However, strong P-Cygni
profiles obscured by \Htwo\ absorption or nebular emission are rare
in the far-UV spectra of our sample.   Some of these are more reliable than
others: \Htwo\ absorption makes the \CIII\ \singlet 977 measurement
questionable if it is weak, and, if strong, the blue edge of its
trough is obscured by \Lyg\ airglow.  \Lyb\ airglow and \Htwo\
absorption similarly affect the \OVI\ doublet.  Furthermore, the
components of the \OVI\ doublet are not optimal estimators of \vinf\
because \OVI\ typically forms deep in the wind.  We usually used \CIV\
\doublet 1548-51 for our first guess of \vinf, and adjusted this value
to match the wind lines.  Our final derived velocities will be
presented in \S~\ref{sec:results} and listed in Table~\ref{tab:mod_param}.

The scarcity of strong wind features of these objects poses some
difficulty in their modeling.  Typically, one determines \Teff\ by the
ionization balance of the CNO elements.  
For these objects, we typically use \CIII/\CIV\ and/or \OIV/\OV/\OVI\
to constrain \Teff.  It is desirable to use diagnostics from many
elements to ensure consistency.  The presence or absence of some
features offer further constraints.  For example, for mass-loss rates
in the regime of our stars, \SiIV\ \doublet 1394,1402
disappears at temperatures $\Teff \gtrsim 50$~kK, \SIV\ \doublet 1062,
1075 at $\sim$45~kK, and \PV\ \doublet\ 1118-28 at $\sim$60~kK.
We generally use all wind lines to constrain \Mdot.

We adopted the following method.  We compared the observed spectra
with the theoretical spectra of our model grids to determine the
stellar parameters \Teff\ and \Rtrans.  far-UV and UV flux levels were
used to set \Rstar\ (since the distance is known), and from that
\Mdot\ and the $L$ can be derived.  If necessary, models with adjusted
abundances are calculated.  Uncertainties in the quoted parameters
reflect the range that the parameters must be varied to fit the
diagnostics.

\subsubsection{Stellar Modeling Results}\label{sec:results}

 From our spectral fitting process we determined the
distance-independent parameters \Teff, \Rtrans, and \vinf.  Once a
distance is adopted, \Rstar\ was derived by scaling the model flux to
the observations, and then $L$ and \Mdot\ were determined.  Our
derived central star parameters are summarized in
Table~\ref{tab:mod_param}, along with masses estimated by comparing
our parameters to the evolutionary tracks of \citet{vassiliadis:94},
and the resulting gravities.  We list our smooth (unclumped) mass-loss
rates $\Mdots$ (corresponding to $f=1.0$), and a clumped rate $\Mdotc$
(corresponding to $f=0.1$).  We have presented our unclumped models in
the figures, but as noted in \S~\ref{sec:results}, the clumped rates
are also generally consistent with the observations. We also list the
ratio of the luminosity to the Eddington luminosity ($\Gamma$), and
indicate which model abundances were used for each object --- SMP
LMC~61 was fit using He/C/O rich abundances typical of ``H-deficient''
(H-poor) objects, and the others with normal, LMC abundances (expected
for ``H-rich'' CSPN).  Noticeable trends that go with increasing
effective temperatures are higher terminal velocities, smaller radii,
and lower mass-loss rates.  These are generally expected from our
current ideas about CSPN evolution --- as the object loses mass as it
evolves along the constant-luminosity portion of the HRD, it contracts
and gets hotter (\eg, \citealp{acker:03}).

Also listed in Table~\ref{tab:mod_param} are reddenings determined from
fitting the multiple components to the far-UV/UV spectra (these are
generally in good agreement with those derived from the Balmer
decrement, shown in Table~\ref{tab:neb}).

We now discuss the modeling results for our sample objects.

\begin{figure}
\begin{center}
\epsscale{1.0}
%\begin{turn}{90}
\rotatebox{0}{
\plotone{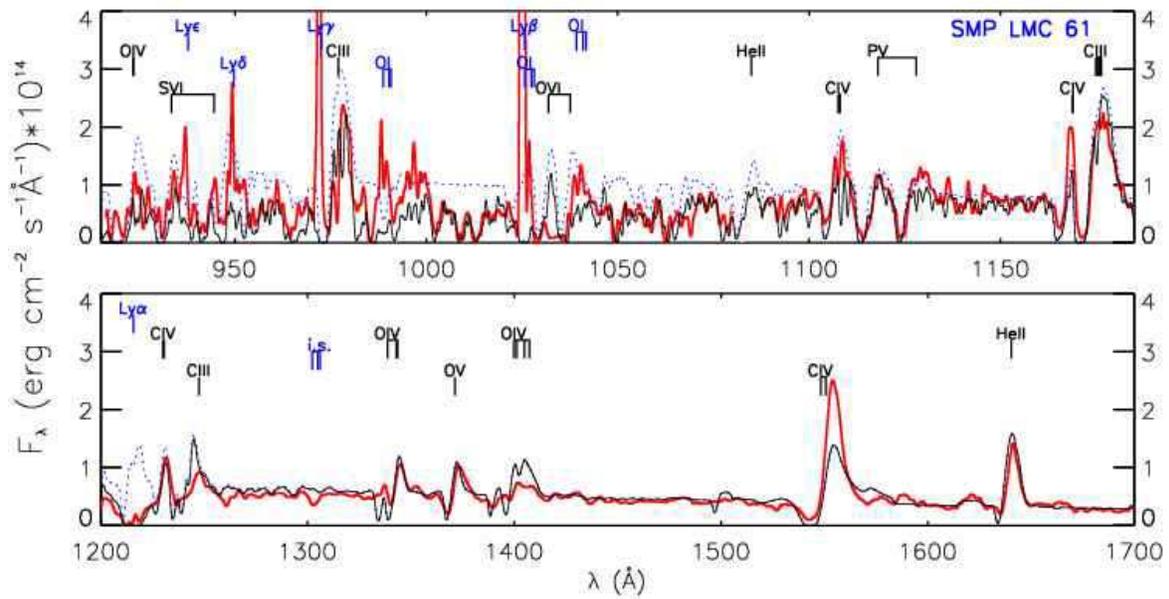}}
%\end{turn}
\caption{SMP LMC~61: The observations are shown (red/gray) along with
  our stellar model, with and without our hydrogen absorption models
  applied (black solid and blue dotted, respectively).  The far-UV
  spectra have been convolved with a 0.6~\AA\ Gaussian for clarity.  }
\label{fig:lmc61mod}
\end{center}
\end{figure}

\noindent \emph{SMP LMC~61 (Fig.~\ref{fig:lmc61mod}):} This object was
classified as a [WC4/5] by \citealp{monk:88} based on its optical
spectrum.  Our model, with $\Teff\simeq70$~kK, $\Mdot \simeq
4.7\E{-7}$~\Msunyr, and $\vinf \simeq 1300$~\kms\ and the He/C/O-rich
(H-deficient) abundances discussed in \S~\ref{sec:abundances} fit most
diagnostics satisfactorily, with the exceptions discussed below.  Our
final abundances for this star are listed in
Table~\ref{tab:lmc61abund}.

Our model (which is unclumped) produces a \CIV\ \doublet 1548-51
weaker than the observed one.  \citet{crowther:02} had similar
problems in their analysis of Galactic WC4 stars. They found the radial velocity
of the \CIV\ profile to be sensitive to clumping in the wind and
achieved better fits with filling factors of $0.1 \le f \le 0.01$.  We have
computed models with $f = 0.1$ and 0.01 to test the effects of
clumping for this [WC] object, and find \CIV\ to also be sensitive to
this parameter.  While the fit of this feature is improved, it is
still not fit satisfactorily in either case.

The \CIII\ \singlet 1247 feature may be blended with \NV\ 1238-43.
Our model, which assumed a nitrogen mass fraction of 1\%, produced a
nitrogen emission which is too strong (presented in the figure).  A
model with no nitrogen (typical of an atmosphere of a massive WC star)
produces better agreement with the observations, although it cannot be
definitively said the nitrogen abundance is zero.

Finally, with an iron abundance 0.4 the solar value, the \FeVI\
spectrum (spanning 1250--1350~\AA) was too strong.  In cooler models,
the \FeV\ spectrum is too strong (appearing between 1350--1500~\AA).
Hotter models produced unobserved \FeVII\ and \FeVIII\ features in the
UV and far-UV, respectively (see \citealp{herald:04b} for a more detailed
discussion of using the iron spectrum as a diagnostic for CSPN).
Smaller mass-loss rates fail to match the non-iron diagnostics.  We
were able to match the iron spectrum by lowering its abundance to
1/10th the LMC metallicity (\XFe = 5.44\E{-5} by mass).  The other
diagnostics are not sensitive to the iron abundance, and removing iron
from the model wind has no significant impact on the synthetic
spectrum (except for \OV\ \singlet 1371, which lies among the iron
spectrum).  This implications of this iron deficiency will be
discussed more in \S~\ref{sec:discussion}.

\citet{bianchi:97} modeled the central star of LMC~61 using a
blackbody distribution to determine parameters of $\Teff=60$~kK,
$\Rstar=0.95$~\Rsun, and $L \simeq 10^{4}$~\Lsun, and $\Mdot =
7\E{-7}$~\Msunyr\ from a wind-line analysis.  However, they had
difficulty in modeling all parts of the spectrum simultaneously
without invoking an additional component (a cooler companion).  They
relied on FOS data, which have flux levels in excess of the
more recent STIS spectrum which we utilize (the STIS and FUSE levels are in
good agreement).

More recently, \citet{stasinska:04} modeled the stellar UV/optical
spectra LMC~61 in conjunction with photoionization models of the
nebula, and determined the central stars parameters: $\Tstar=88$~kK,
$\Rstar=0.42$~\Rsun, $\log{L} = 3.96$~\Lsun, $\log{\Mdot} =
-6.12$~\Msunyr, with He/C/O = 0.45/0.52/0.03.  Their mass-loss rate
agree with ours (within the uncertainties), but they find a higher
luminosity and temperature.  However, they were not able to
simultaneously fit the observed spectrum and nebular ionizing fluxes,
noting that the nebular analysis requires a central star luminosity of
$\log{L} = 3.66$~\Lsun\ (in line with our findings).  The higher
temperature they derived appears to be due to choice of diagnostics.
For example, they fit \CIV\ \doublet 1548-51 well (which we do not)
but not \CIII\ 1175 (which we do).  They, too, found that a sub-LMC
iron abundance was needed to match the spectrum of this object.

\begin{figure}
\begin{center}
\epsscale{1.0}
%\begin{turn}{90}
\rotatebox{0}{
\plotone{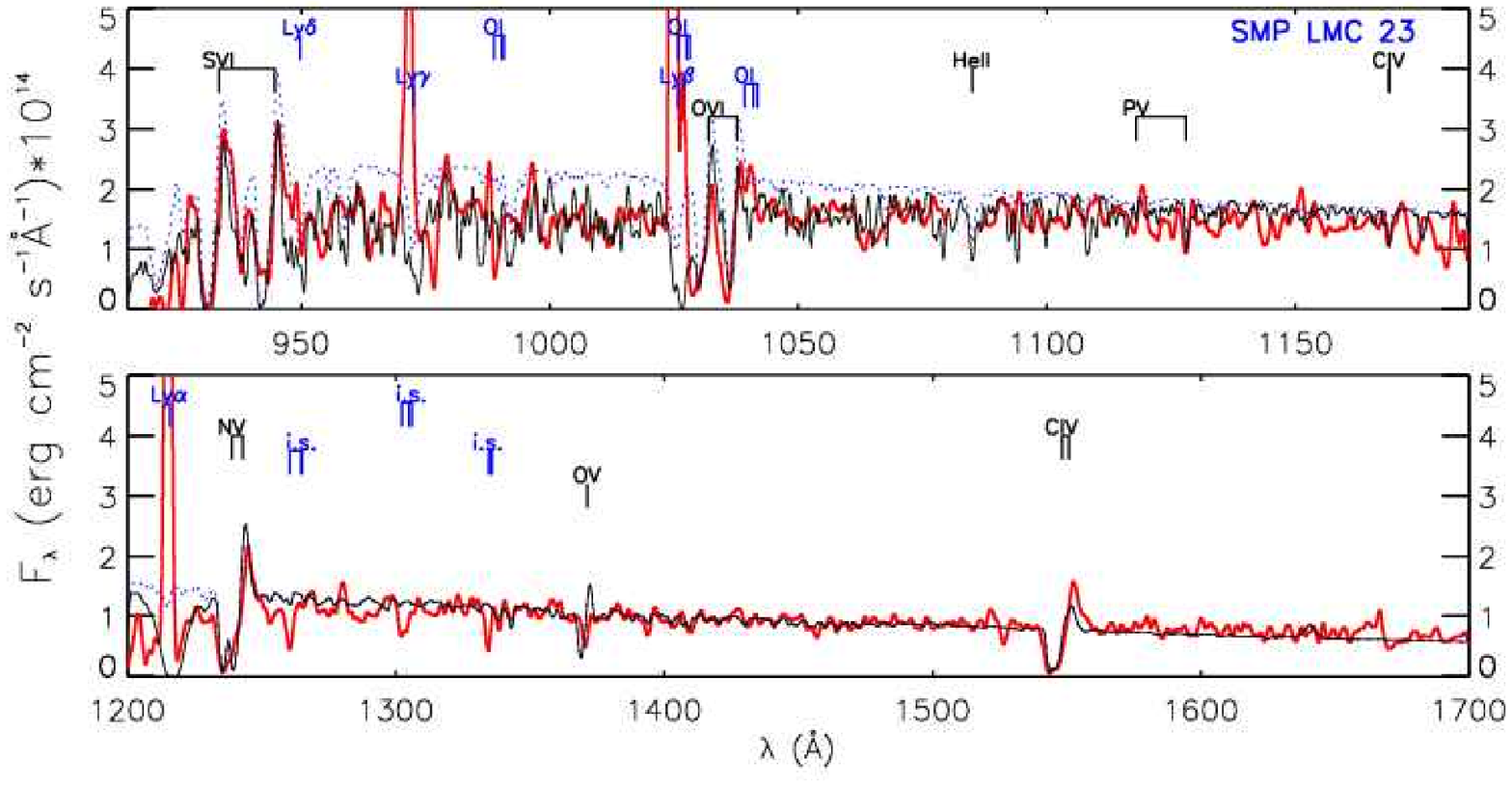}}
%\end{turn}
\caption{SMP LMC~23: Description follows that of Fig.~\ref{fig:lmc61mod}.
 }
\label{fig:lmc23mod}
\end{center}
\end{figure}

\begin{figure}
\begin{center}
\epsscale{1.0}
%\begin{turn}{90}
\rotatebox{0}{
\plotone{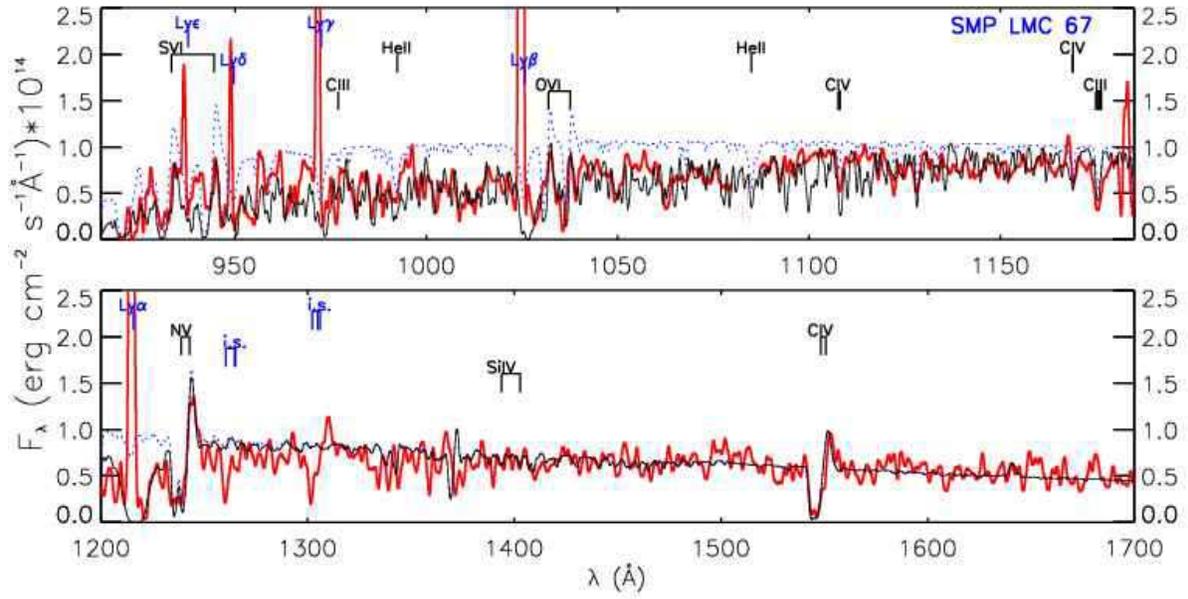}}
%\end{turn}
\caption{SMP LMC~67: Description follows that of
  Fig.~\ref{fig:lmc61mod}. The theoretical nebular continuum
  (\S~\ref{sec:nebcont}) has been
  subtracted from the UV observations.}
\label{fig:lmc67mod}
\end{center}
\end{figure}

\noindent \emph{SMP LMC~23 and 67 (Figs.~\ref{fig:lmc23mod} and
  \ref{fig:lmc67mod}):} As noted before, the spectra of SMP
LMC~23 and SMP LMC~67 are similar (except for \CIII\ \singlet 1175),
and we derive similar parameters: $\Teff=60$~kK,
$\Mdot=3.8\E{-8}$~\Msunyr\ and $\Teff=55$~kK,
$\Mdot=4.5\E{-8}$~\Msunyr, respectively.  We fit both adequately using
models of normal LMC abundances (\ie, ``H-rich'').  SMP LMC~67 is a
bit hotter, as evidenced by stronger \OVI\ and lack of \CIII\ \singlet
1175.  The far-UV flux of both objects suffer significant absorption due
to hot \Htwo, which can be appreciated by comparing the models before
and after the absorption effects are applied in the figures.  In the
UV, the nebular continuum contributes a bit to the flux for SMP LMC~67.
For these objects, clumped models with filling factors of $f=0.1$
showed little change in the emergent spectra, with only \OVI\ 1032-38
weakening slightly.

\citet{bianchi:97} modeled the central star of SMP LMC~67 using a
blackbody distribution and determined $\Teff=45$~kK,
$\Rstar=1.06$~\Rsun, and $L = 4157$~\Lsun\ (versus our values of
$\Rstar=0.80$, $L \simeq 5000$~\Lsun).

\begin{figure}
\begin{center}
\epsscale{1.0}
%\begin{turn}{90}
\rotatebox{0}{
\plotone{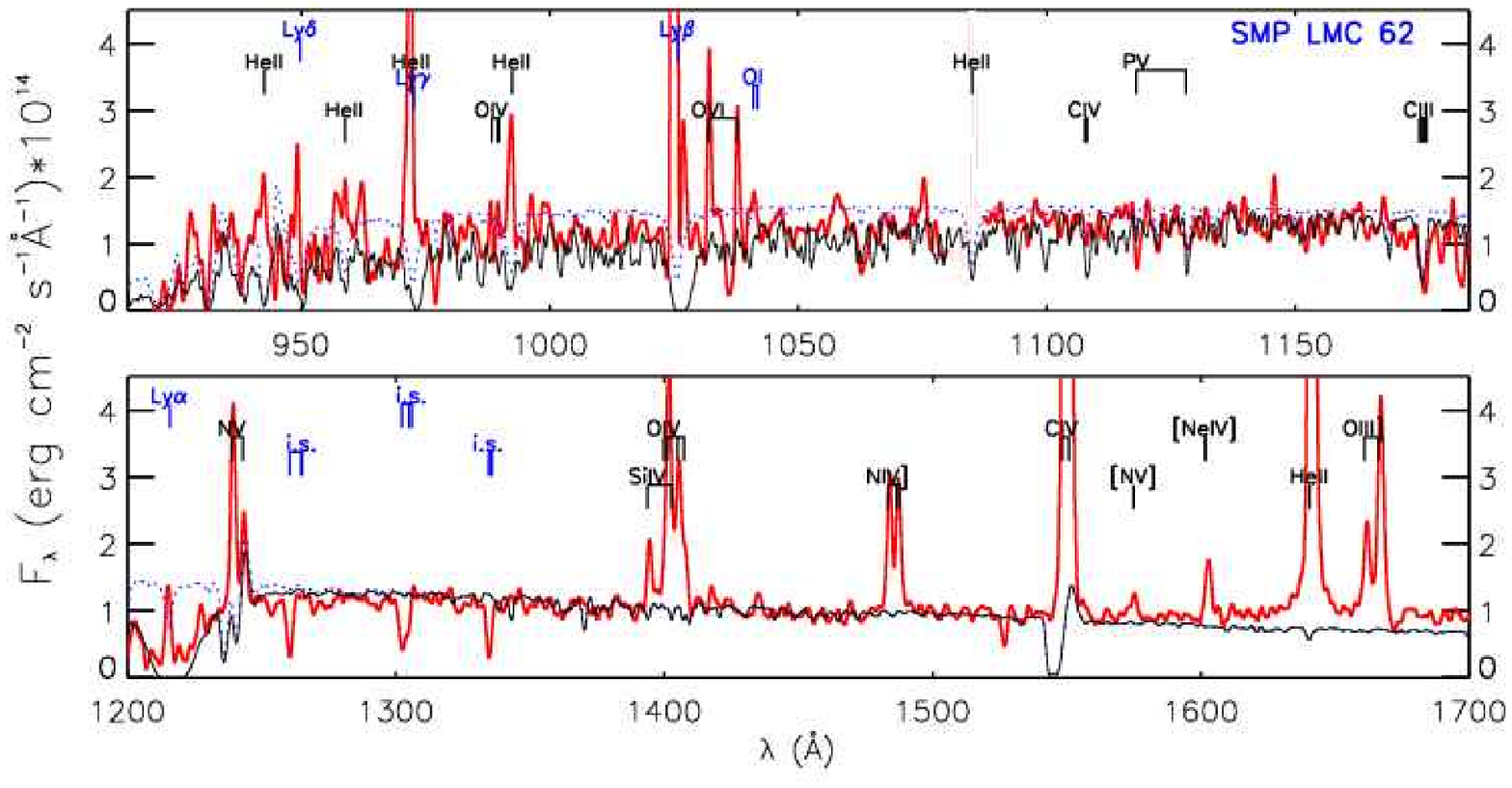}}
%\end{turn}
\caption{SMP LMC~62: Description follows that of Fig.~\ref{fig:lmc61mod}.
}
\label{fig:lmc62mod}
\end{center}
\end{figure}

\noindent \emph{SMP LMC~62 (Fig.~\ref{fig:lmc62mod}):} The FUSE
spectrum of SMP LMC~62 is mostly the fingerprint of hot ($T \gtrsim
3000$) \Htwo.  Even what appears to be a P-Cygni profile for the \OVI\
doublet might be an artifact of the \Htwo\ absorption.  Once the
\Htwo\ is accounted for, there are no obvious strong wind features
detectable in the far-UV.  \SVI\ \doublet 933-44 is not obviously
present.  \CIII\ \singlet 1175 appears in absorption, but it may have
an interstellar contribution (\citealp{danforth:02} found \CIII\
absorption associated with LMC material along many of the LMC
sight-lines presented in their comprehensive atlas).  \CIV\ \singlet
1107 and \PV\ \doublet 1118-28 seem to be present, but because of the
\Htwo\ absorptions which completely blanket the FUSE range, a
definitive statement is not possible.  The nebular emission features in the
FOS range make it difficult to discern any stellar features except
for, perhaps, \NV\ \doublet 1238-43.  Thus, this star has fewer
diagnostics to work with than the others of our sample. We have
assumed H-rich abundances for this object, however, due to the lack of
stellar features, abundances cannot be firmly constrained.  \CIII\
\singlet 1175 and the absence of strong \SVI\ \doublet 933-44 place a
rough upper limit on \Teff\ of $\lesssim 55$~kK.  The presence of
fairly strong \NV\ \singlet 1240 requires $\Teff \gtrsim 40$~kK, and
imposes a rough lower limit to \Mdot.  Our model adequately fits the
emission of this feature, but because the nebular emission obscures
the absorption trough, \vinf\ is difficult to constrain.  \PV\ and
\CIII\ appear in absorption, restricting \Mdot.  The absence of \OIV\
\doublet 1339-43 and \OV\ 1371 also places upper limits on \Teff\ and
\Mdot.  Thus, we estimate $\Teff = 45\pm5$~kK, $\Mdot \simeq
2\E{-8}$~\Msunyr, and $\vinf \simeq 1000$~\kms\ for this star.

As discussed in \S~\ref{sec:htwo}, the extremely high temperature
required to fit the \Htwo\ absorptions suggests non-equilibrium
conditions in the nebular environment, which might arise from shocks.
This sight-line to this star also has a relatively high \HI\ column
density [$\logN(\HI) \simeq 21.7$] with respect to the others of our
sample.  The reddening corresponding to this column density for a
typical interstellar dust/gas ratio is much higher than what we
determine from our continuum fits ($\EBMV \simeq 1$ vs. 0.1 --- see
Tables~\ref{tab:hydrogen} and \ref{tab:mod_param}).  This implies that
the majority of \HI\ along the sight-line is probably circumstellar.

Previous photoionization models of the UV nebular spectrum (discussed
in \S~\ref{sec:discussion}) have predicted a significantly higher
temperature for this star than our result (127 vs. 45~kK).  Central
star temperatures of $\Teff > 125$~kK produce nebular \OVI\ \doublet
1032-38 through photoionization \citep{chu:04}, so such a temperature would be
consistent with our observations of this feature.  Our low stellar
temperature can be reconciled with the highly-ionized nebular spectrum
if shocks (suggested by the \Htwo\ features) are responsible for the
nebular ionization.  However, in shock excited gas [\NeV] 3426 is
invariably much weaker than \OVI\ \singlet 1032.  Therefore our
observations seem to rule out shocks as the source of the [Ne V]
emission.  In contrast, photoionization models of high excitation PN
\citep{otte:04} predict a [Ne V] 3426 to O VI 1032 ratio of about 7,
which is similar to what we observe.

Test stellar atmosphere models at the photoionization temperature
($\Teff = 127$~kK) do not reproduce \CIII\ \singlet 1175 or \PV\
\doublet 1118-28 (although the former could be interstellar and the
latter cannot be clearly seen due to \Htwo\ absorptions).  For this
temperature, the observed far-UV flux levels imply $\Rstar \simeq
0.5$~\Rsun\ and $\log{L} \simeq 4.7$~\Lsun, an extremely luminous CSPN
(most LMC CSPN have $\log{L} \lesssim 3.9$~\Lsun, \eg,
\citealp{dopita:91a,dopita:91b}).  This luminosity correspond to
current mass of $\sim$1.2~\Msun\ (according to the tracks of
\citealp{paczynski:70}) and is comparable to that of LMC N~66 (SMP
LMC~83), a WN-type object which displayed a dramatic outburst in 1993
and the nature of which remains unclear \citep{hamann:03}.  Thus, the
results of our stellar modeling are not consistent with parameters
derived from nebular analysis, and the discrepancy is not clarified by
the present data.  CSPN with winds and high temperatures such as the
photoionization temperature of this object typically belong to a class
of objects termed ``\OVI\ PN nuclei'', which exhibit emission lines of
\OVI\ \doublet 3811-34 and, often, \CIV\ \doublet 5801-12.  Thus
optical spectroscopy of the central star could help resolve the
discrepancy.

\begin{figure}[htbp]
\begin{center}
\epsscale{1.0}
%\begin{turn}{90}
\rotatebox{0}{
\plotone{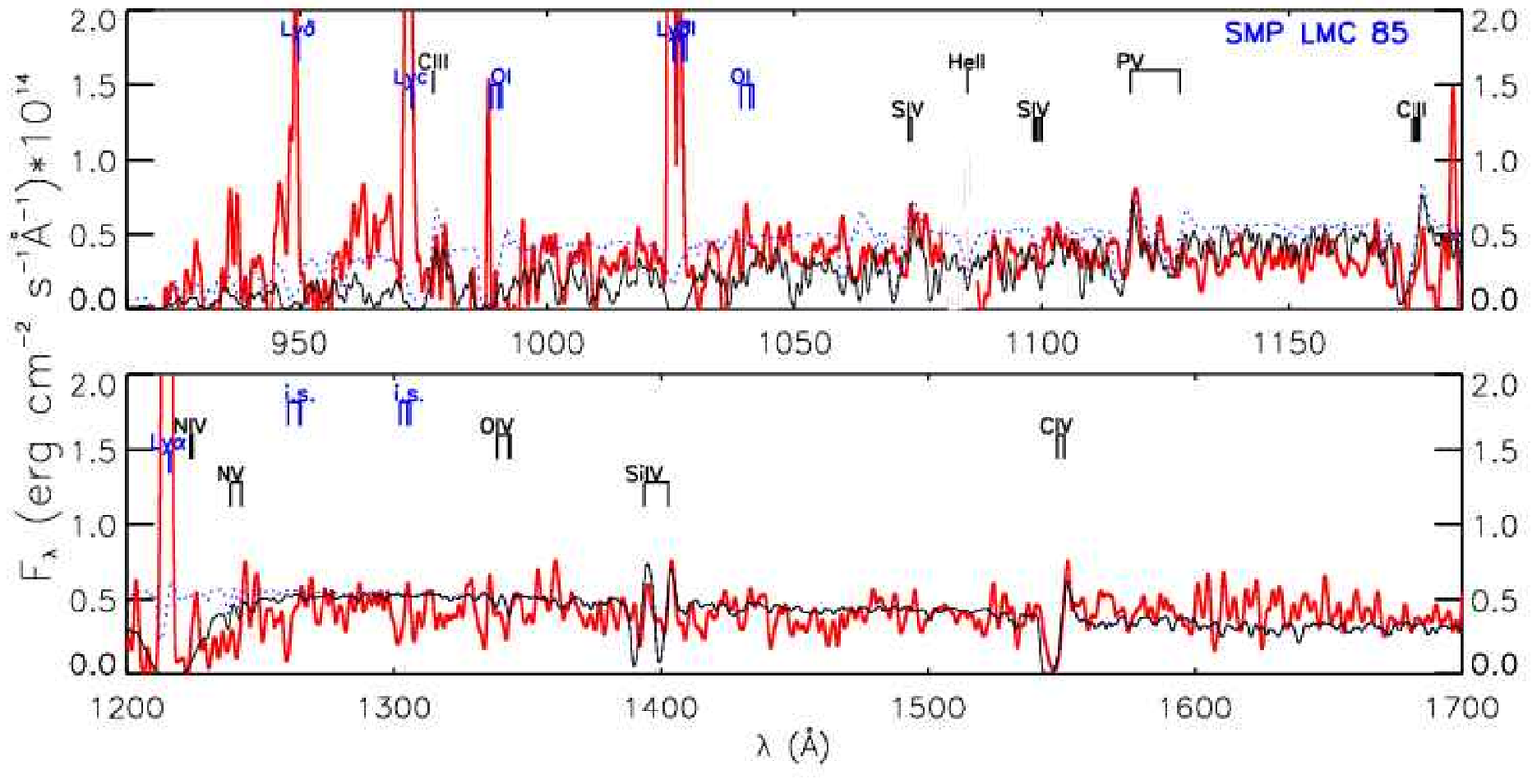}}
%\end{turn}
\caption{SMP LMC~85: Description follows that of Fig.~\ref{fig:lmc61mod}.
}
\label{fig:lmc85mod}
\end{center}
\end{figure}

\noindent \emph{SMP LMC~85 (Fig.~\ref{fig:lmc85mod}):} This is
one of the dimmer objects in our sample (it is more reddened). The far-UV
spectrum of this object is somewhat contaminated with airglow lines,
and features at shorter wavelengths should be viewed with caution due
to the small effective area of the FUSE SiC detectors.  However, one
can still recognize \CIII\ \singlet 977, \SIV\ \singlet 1072, \PV\
\doublet 1118-28 and perhaps \OVI\ \doublet\ 1032-38 and \CIII\
\singlet 1175 in the far-UV.  In the UV, the \SiIV\ and \CIV\ doublets
are present, and perhaps \OIV\ \doublet 1339-43 (the FOS data are noisy also).
The \PV, \SIV, and \SiIV\ lines indicate a cooler object with a
lower terminal velocity.  We derive $\Teff = 40\pm2$, $\Mdot \simeq
4-16\E{-8}$~\Msunyr, $\vinf \simeq 700$~\kms, and are able to
fit its spectrum adequately using LMC abundances.

\begin{figure}
\begin{center}
\epsscale{1.0}
%\begin{turn}{90}
\rotatebox{0}{
\plotone{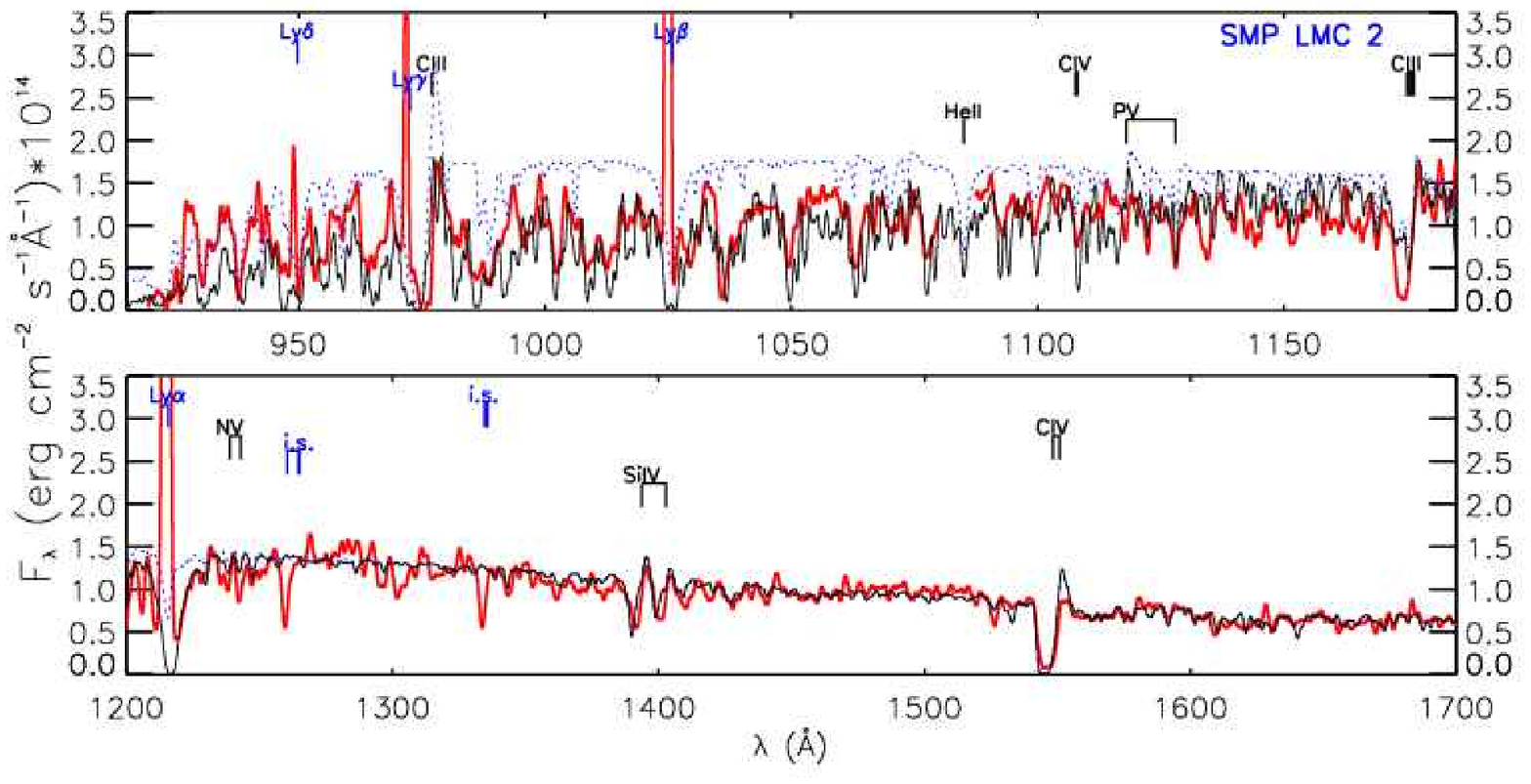}}
%\end{turn}
\caption{SMP LMC~2: Description follows that of Fig.~\ref{fig:lmc61mod}.
}
\label{fig:lmc2mod}
\end{center}
\end{figure}

\noindent \emph{SMP LMC~2 (Fig.~\ref{fig:lmc2mod}):} This object has
strong \CIII\ signatures in the far-UV (both \singlet 977 and \singlet
1175).  Once the (hot) \Htwo\ absorption is taken into account, it
appears the \OVI\ doublet is not present.  The UV spectrum shows
modest \NV, \SiIV\ and \CIV\ doublets.  We derive $\Teff \simeq
38$~kK, $\Mdot \simeq 3\E{-8}$~\Msunyr, and $\vinf \simeq 700$~\kms\
using LMC abundances.  In our unclumped model, \PV\ \doublet 1118-28 is a bit
strong.  A clumped model with $f=0.1$ resulted in the \PV\ feature to
weaken enough to agree with the observations, while leaving the other
wind features unchanged significantly (we present the unclumped model
in the figure, for consistency with the other objects).

\begin{figure}
\begin{center}
\epsscale{1.0}
%\begin{turn}{90}
\rotatebox{0}{
\plotone{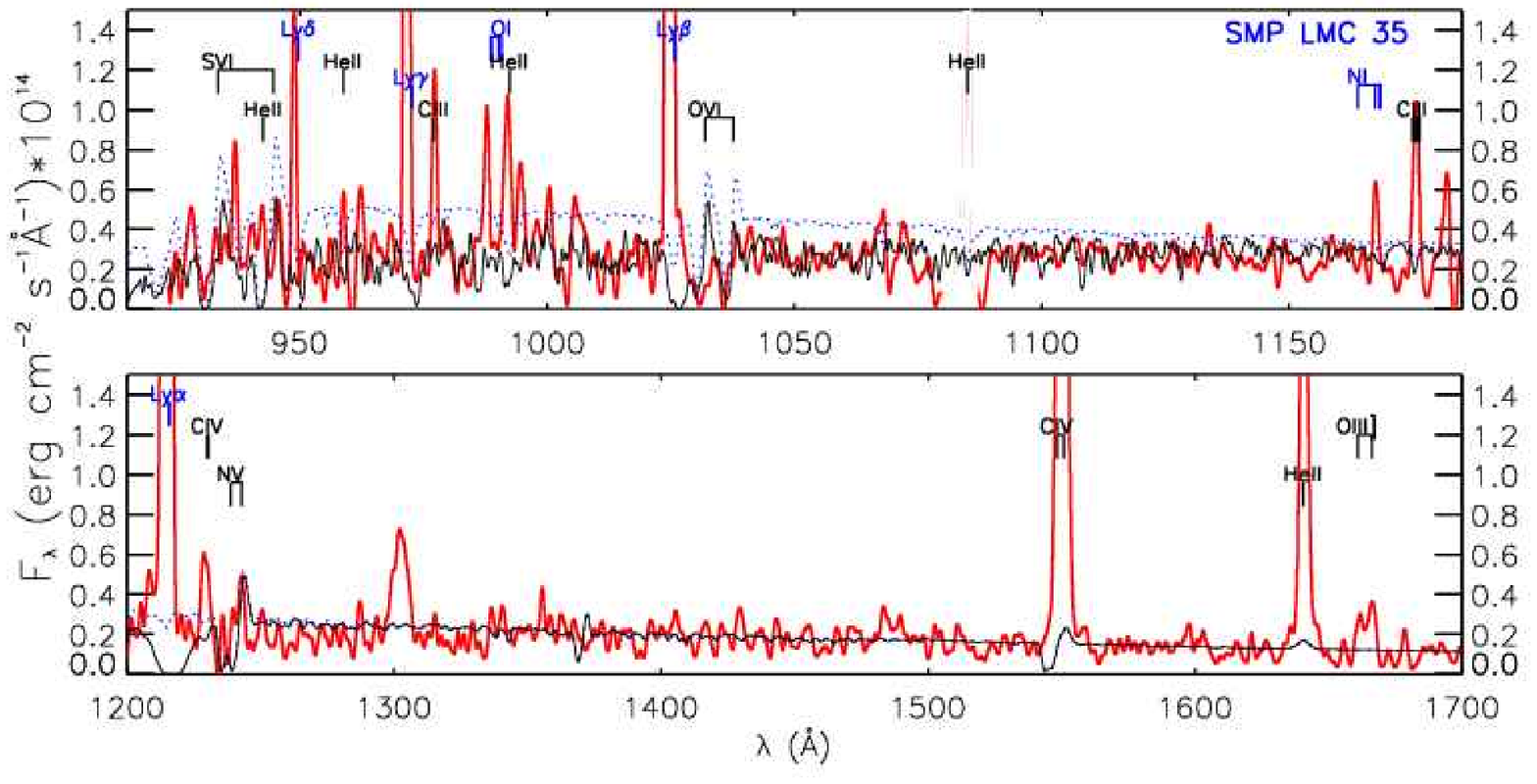}}
%\end{turn}
\caption{SMP LMC~35: Description follows that of Fig.~\ref{fig:lmc61mod}.
}
\label{fig:lmc35mod}
\end{center}
\end{figure}

\noindent \emph{SMP LMC~35 (Fig.~\ref{fig:lmc35mod}):} The dimmest
object of our sample, LMC~35 is shown in Fig.~\ref{fig:lmc35mod}.  The
data has the lowest S/N, and has been heavily smoothed.  Many
features, especially at the shorter wavelengths covered by FUSE, are
dubious.  Given the poor quality of the data, we did not attempt a
rigorous fit of the stellar spectra.  We present a model of $\Teff =
50$~kK, $\Mdot=1\E{-8}$~\Msunyr, and $\vinf = 1000$~\kms\ that is
produces an acceptable fit.
The data are, however, good enough to see that hot \Htwo\ lies along
this objects sight-line.

\section{DISCUSSION}\label{sec:discussion}

Our sample PN are compact, and thus relatively young.  Our derived
effective temperatures ($38 \lesssim \Teff \lesssim 70$~kK) are on the
cooler side for CSPN.  The sample also show a small spread in
luminosity.  For comparison, temperatures derived from
photoionization models
\citep{dopita:91a,dopita:91b,dopita:97,vassiliadis:98} are listed in
Table~\ref{tab:photo}.  Our derived temperatures are in good agreement
with those of the photoionization models, with the exception of SMP
LMC~62 (and SMP LMC~35, which we did not fit).  For SMP LMC~62, we derive a
significantly lower temperature than the photoionization temperature
(45 vs. 127~kK --- discussed in \S~\ref{sec:results}).

Comparing our temperatures and luminosities with the $z=0.008$
evolutionary tracks of \citet{vassiliadis:94} (appropriate for the
metallicity of the LMC), we find our sample luminosities are a bit low
for the H-burning tracks, and fall more naturally on their He-burning
tracks.  Comparing with the latter, our hotter objects (\ie, $\Teff
\gtrsim 45$~kK) fall between the initial-final mass tracks of
($M_i,M_f$) = (2.0,0.669) and (1.5,0.626)~\Msun, with evolutionary
ages of 1-3~kyr.  The cooler objects lie within the (1.0,0.578) and
(0.95,0.554)~\Msun\ tracks, also with evolutionary ages of 1-3~kyr.
These evolutionary ages, $\tau_{evol}$, are listed in
Table~\ref{tab:photo} along with the dynamical ages calculated from
$\tau_{dyn} = r_{neb}/v_{exp}$ using the values from
Table~\ref{tab:neb}.  Note that 
the kinematic age is expected to be a lower limit to the post-AGB age,
as the nebular expansion is thought to increase during the early
post-AGB phase and then level off as the nucleus fades (see
\citealp{sabbadin:84}, \citealp{bianchi:92} and references therein).
Generally, the dynamic and evolutionary ages are in good agreement,
not differing by more than a factor of two.  It is not surprising that
most of the objects are on the massive side.  This is because the
brightest known objects were chosen for the FUSE program since their
fluxes are at the lower limits of the FUSE sensitivity.

The low mass-loss rates of these objects and their lack of diagnostics
make the abundances difficult to constrain.  For most objects, we are
able to fit their spectra using normal LMC abundances (corresponding
to H-rich CSPN).  H-rich CSPN can be either H- or He-burning, but
comparison with the evolutionary tracks indicate that these objects
are most probably He-burners.  

SMP LMC~61, the [WC4] star, is fit using He/C/O enriched abundances
that only He-burners are expected to display.  We also find an iron
deficiency for this object (as do \citealp{stasinska:04}).
\citet{werner:02} and \citet{herald:04b} have found iron deficiencies
in Galactic PG~1159-[WC] stars (these stars are thought to represent a
phase where the CSPN is transitioning between a [WC] star and PG~1159
star, an entry point onto the WD cooling sequence).  \citet{miksa:02}
have also found iron deficiency in a large sample of (Galactic)
PG~1159 stars, the supposed descendants of these transitional objects.
Iron deficiencies in these objects may result when material in the
He-intershell is exposed to $s$-process nucleosynthesis during a
thermally pulsating AGB or post-AGB phase \citep{lugaro:03,herwig:03}.
Iron deficiency in [WC] stars is very interesting from the standpoint
of radiation-driven winds.  For massive WR stars, the opacity of the
wind from iron and iron-group elements is thought to play a crucial
role in initiating the driving of the wind
\citep{lamers:02,crowther:02}.  The iron opacity of these optically
thick line driven winds plays an important role in determining where
the sonic point occurs, which in turn dictates the characteristics of
the wind (\eg, \Mdot) \citep{lamers:02}.  Wind density is the primary
discriminator between the (massive) WC4-7 subtypes
\citep{crowther:02}.  In contrast to massive WC stars, there is a
known lack of [WC5-7] subtypes among CSPN, which is not understood (\eg,
\citealp{crowther:02} and references therein).  If [WC] winds are
similarly sensitive to the iron opacity, an iron deficiency in these
winds might be related to the dearth of [WC5-7] subtypes among CSPN.

\citet{vink:01} presented mass-loss rate predictions for
\emph{massive} O and B stars of different metallicities.  Their models
yield mass-loss rates for given stellar parameters, taking into
account opacity shifts in the wind due to different ionization
structures at different temperatures (assuming ``normal'', \ie,
non-enriched, abundances).  We tested their relation for our
\emph{low-mass} objects using our derived stellar parameters (\Teff,
$L$, \vinf, and \logg) from Table~\ref{tab:mod_param} with our adopted
metallicity of $z=0.4\Zsun$ (note that SMP LMC~61 has chemically processed
atmosphere, so the Vink prescription does not apply in this case).
The results (listed in Table~\ref{tab:photo}) show, on average, the
predicted mass-loss rates to be $\sim 3$ times
lower than our derived (smooth-wind) rates.  However, a clumping
factor of $f=0.1$ (which is consistent with observations in most cases
and produce better fits in some ---
\S~\ref{sec:results}) would reduce our mass-loss rates sufficiently to
bring them into rough agreement with the \citet{vink:01} values.  

Also listed in Table~\ref{tab:photo} for our objects are the ratios of
the wind momentum flux to the radiative momentum flux (the
``performance numbers''), defined as $\eta = \vinf \Mdot c/L$
\citep{springmann:94}, calculated using the parameters in
Table~\ref{tab:mod_param}.  A performance number of unity corresponds
to the case where each photon, on average, scatters once in the wind
(the ``single scattering limit'').  A performance number of $>1$
indicates photons are scattering multiple times, requiring a higher
wind opacity.  For all the objects of our sample, $\eta \lesssim 1$
except for SMP LMC~61, which has $\eta \simeq 7$.  There are
indications that the wind of the latter object is clumped
(\ref{sec:results}) which would reduce \Mdot.  But the wind would have
to be highly clumped ($f < 0.01$) to reduce $\eta$ below unity (it was
not necessary to invoke clumping to fit the profiles of the
LMC-abundance objects).  The larger performance number for the [WC]
object reflects the higher opacity of its chemically enriched wind,
which makes the wind more efficient at capturing radiation momentum.

\citet{acker:03} analyzed and classified 42 Galactic [WR] objects,
which included 7 [WC4] stars.  Their average effective temperature was
the same as for our LMC [WC4] star, SMP LMC~61 (70~kK).  But their
average terminal velocity was almost twice as high (2300
vs. 1300~\kms).  As the [WR] evolves through the constant-$L$ phase,
$\vinf/\Mdot$ is roughly constant and \vinf\ correlates with \Teff\
(\eg, \citealp{leuenhagen:98} and references therein).  Thus the lower
\vinf\ of our LMC object with respect to Galactic objects of similar
\Teff\ and $L$ probably is an effect of the lower LMC
metallicity.

To further investigate the effects of metallicity on the radiative driving, we
computed the \emph{modified wind momentum} $\Pi \equiv
\Mdot\vinf\Rstar^{0.5}$ for our sample CSPN, also listed in
Table~\ref{tab:photo}.  Radiative wind driven theory predicts $\Pi
\propto Z^{1.0}$.    For Galactic (massive) O and B-stars, there is a
clear correlation between $L$ and $\Pi$.  \citet{lamers:96} have shown
that Magellanic Cloud O and B stars of equivalent luminosities do
indeed tend to have lower modified wind momenta, although not quite
as low as predicted by theory.  

\begin{figure}
\begin{center}
\epsscale{.8}
%\begin{turn}{90}
\rotatebox{270}{
\plotone{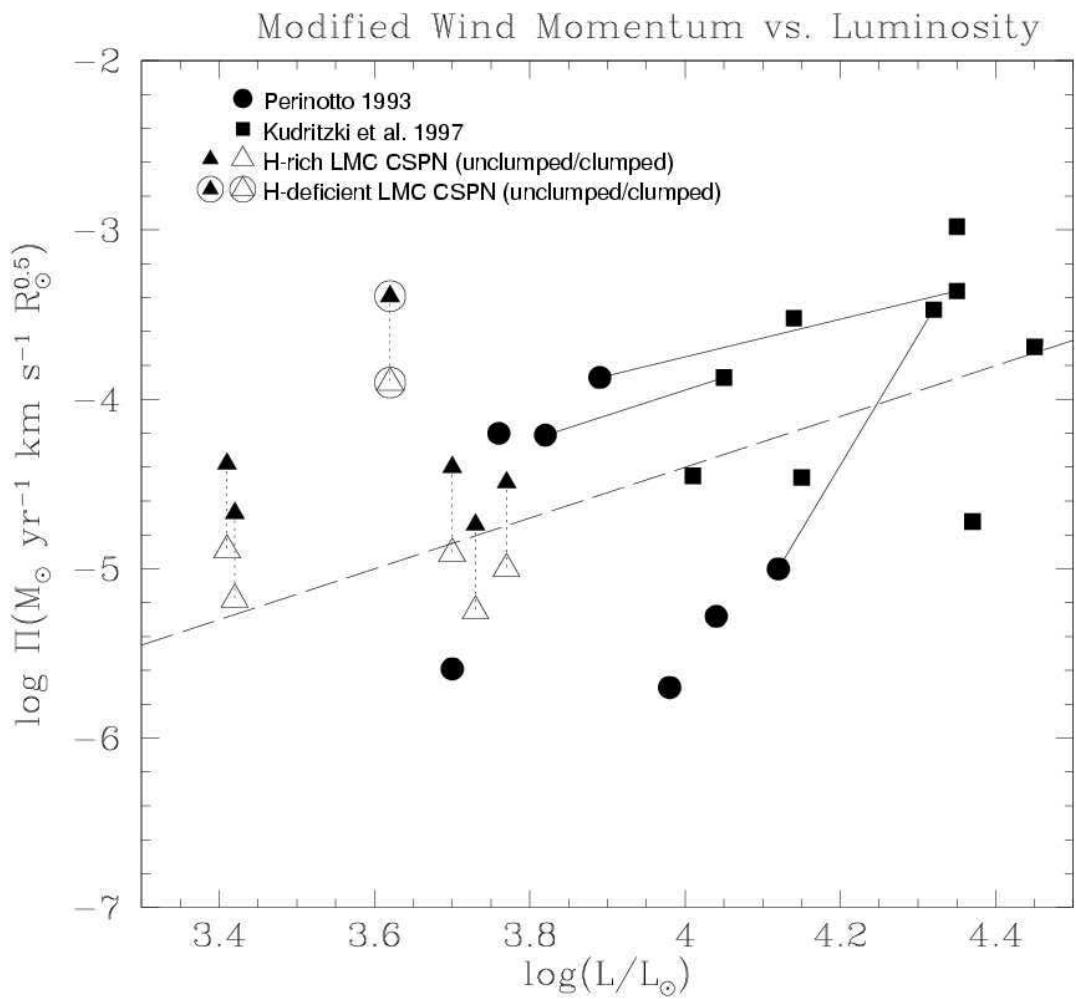}}
%\end{turn}
\caption{Luminosity vs. Modified Wind Momentum: The Galactic samples
  of \citet{perinotto:93} and \citet{kudritzki:97} are shown (solid
  circles and squares, respectively).  Solid lines join stars
  appearing in both Galactic samples. We also show our LMC sample,
  with parameters corresponding to smooth and clumped winds (filled
  and unfilled triangles, connected by dotted line).  SMP LMC~61, the
  [WC] star, is denoted by the circled triangles.  (unfilled
  triangle).  The luminosity-modified wind momentum relation for
  Galactic O-stars, extrapolated to CSPN luminosities, is shown by the
  dashed line (from \citealp{tinkler:02}).  }
\label{fig:wind}
\end{center}
\end{figure}

\citet{tinkler:02} calculated $\Pi$ for the H-rich CSPN samples of
\citet{kudritzki:97} and \citet{perinotto:93}, and extrapolated the
luminosity-wind momentum relation for Galactic O-stars down to CSPN
luminosities.  They found the H-rich CSPN to be clustered into two
groups, with higher and lower $\Pi$s than predicted by the relation,
respectively, but with about the same slope.  In Fig.~\ref{fig:wind},
we plot these Milky Way samples with our objects, with both our
unclumped mass-loss rates, and mass-loss rates corresponding to a clumping
factor of $f=0.1$.  Our objects extend the high-$\Pi$ group to lower
luminosities, but basically follow the same relation.  In other words,
they do not show systematically lower wind momenta as predicted for
their lower metallicity.  Possible explanations include that the wind
momentum-luminosity for massive OB stars is not valid for lower CSPN
luminosities and/or our luminosities are underestimated from our lack
of knowledge of $\logg$ (see \S~\ref{sec:gravity}).  To expand on the
former, \citet{tinkler:02} attempted to reconcile the differences
between the \citet{kudritzki:97} and \citet{perinotto:93} samples by
scaling their parameters to make them consistent (although, they
admit, not necessarily correct).  Investigating the wind
momentum-luminosity relationship with this rescaled set, they
concluded that, for CSPN, the wind momentum seemed to depend more on
\Teff\ than on $L$.  If that is correct, metallicity may have less of
an impact on the evolution of low/intermediate mass stars than on
massive ones.  However, it should be noted that all three samples
(\citealp{perinotto:93}, \citealp{kudritzki:97}, this work) were
analyzed using different methods that are not consistent (note the
difference of parameters for NGC~6826 and IC~418, which appear in both
Galactic works).  In particular, \citet{tinkler:02} note that the
\citet{kudritzki:97} sample most likely overestimates \Rstar\ (and
hence, \Mdot\ and $L$) due to their reliance on the \Hgamma\ line as a
diagnostic.  The root of the discrepancies seen among the Galactic
samples are also due to distance uncertainties, a problem that does
not affect our LMC sample.

All of our objects have very hot molecular hydrogen ($T \sim 2000$~K)
along their sight-lines, presumably associated with their
circumstellar environment.  \citet{herald:02} found a \Htwo\ gas of
$T\simeq 1250$~K around the (more evolved) Galactic CSPN of Abell~35,
and \citet{herald:04b} found \Htwo\ temperatures of $\sim300$~K for
four other Galactic CSPN.  The higher \Htwo\ temperatures found in our
LMC objects may be because they are compact, thus more likely a shocked
environment (as indicated by SMP LMC~62) and/or because the \Htwo\
is closer to the star.  \Htwo\ may exists in
clumps, shielded from the intense UV radiation fields by neutral and
ionized hydrogen, as appears to be the case in the Helix nebula
\citep{speck:02}.  \citet{speck:02} suggest that these clumps may form
after the onset of the PN phase, arising from Rayleigh-Taylor
instabilities at either the ionization front or the fast wind shock
front.  It could be that this happened in the recent enough past
history of these young PN that the \Htwo\ has not returned to its
equilibrium state.
  
As noted in \S~\ref{sec:htwo}, our measured column densities of \HI\
(from the \Lya\ and \Lyb\ profiles) are higher than those determined
by applying a typical interstellar gas-to-dust relation (such as that
of \citealp{bohlin:78}) to our derived reddenings (listed in
Table~\ref{tab:hydrogen}).  Assuming that the difference represents
the \emph{circumstellar} \HI\ column, we calculated the circumstellar
\HI\ mass for a variety of simple shell geometries.  Because the \HI\
column density is higher (a few order of magnitudes) than the \Htwo\
column density, and the typical mass of the ionized gas is negligible,
we can assume that the neutral hydrogen accounts for the difference
between the initial mass (inferred from the evolutionary tracks,
\S~\ref{sec:discussion}) and the mass of the remnant
(Table~\ref{tab:mod_param}).  In all but one case the comparison
suggests the \HI\ to be located in a volume larger than (or outside
of) the ionized shell.  The exceptional case is SMP LMC~62, where the
\HI\ seems confined within a comparable radius to the ionized gas.

\section{CONCLUSIONS}\label{sec:conclusions}

We have analyzed FUSE observations of seven central stars of bright,
compact planetary nebulae in the LMC.  Most objects display definite
wind features, and we determined their stellar parameters using the
stellar atmosphere code CMFGEN to analyze their FUSE far-UV and HST UV
spectra.  We also modeled the nebular continua (which contributes to the
UV flux) and the atomic and molecular hydrogen absorption along the
sight-line (which severely affects the far-UV spectra).

By virtue of their membership of the LMC, the uncertainties in the
distances is small.  This is a great advantage over Galactic CSPN,
where the distance is the largest source of uncertainty in the
analysis.  The objects have a spread of effective temperatures between
35 and 70~kK, with mass-loss rates of $\Mdot \sim 5\E{-8}$~\Msunyr\
(the one [WC] star has a mass-loss rate an order of a magnitude
larger).  Terminal wind velocities generally increase with increasing
effective temperatures, and range between 700---1300~\kms, a factor of
two lower than Milky Way counterparts.  Radii decrease with
increasing temperature, as expected.  Their luminosities are similar ($L \sim
4000$~\Lsun), and the parameters of the CSPN fall on the
He-burning evolutionary tracks of \citet{vassiliadis:94}, from which we infer
post-AGB ages in good agreement with the estimated dynamical ages.

We modeled five of the objects with typical LMC abundances ($z =
0.4$~\Zsun).  For these, we investigated the effects of the
lower LMC metallicity on the wind acceleration by calculating the modified
wind momentum.  Contrary to what is expected based on radiative driven
wind theory, we did not find any substantial deviation from the wind
momentum-luminosity relation that holds for Galactic O
and B stars, and roughly holds for Galactic CSPN stars, but the
scatter is comparable to the expected difference.  Whether this is 
because the wind momentum does not as simply depend on the luminosity
for CSPN as for O and B stars (as suggested by \citealp{tinkler:02})
or because of some other effect is unclear.

The one [WC] star of our sample, SMP LMC~61, was modeled using typical
He/C/O enriched abundances (products of He-burning).  We also
determined a sub-LMC iron abundance (also found by
\citealp{stasinska:04}), perhaps a product of $s$-process
nucleosynthesis during a thermally pulsating AGB or post-AGB phase
\citep{lugaro:03,herwig:03}.  This is particularly interesting, as
iron opacity is thought to play a key role in the driving of WC winds
\citep{lamers:02,crowther:02}. SMP LMC~61 has a wind terminal velocity
about half that of Galactic [WC4] stars from the sample of
\citet{acker:03}.

Our analysis of the far-UV and UV spectra also provided insight into
the circumstellar environment of these objects.  Our measured \HI\
column densities are higher than those predicted by typical
interstellar gas-to-dust relations using our derived reddenings, which
implies a significant amount of circumstellar \HI, presumed to have
once been a part of the progenitor object.  We calculated some simple
shell models which implied most of this material lies outside the
ionized radius of the nebula of these stars, with the exception of SMP
LMC~62, for which it seems to lie within.  This object also displays
nebular \OVI\ \doublet 1032-38, (such features have been observed in
the Galactic CSPN NGC~2371 by \citealp{herald:04b}).

The high resolution FUSE spectra revealed that these
objects have very hot ($T \gtrsim 2000$~K) molecular hydrogen in the
circumstellar environment.  These temperatures may be due to the
proximity of the nebular gas to the star, or perhaps to shocks.

In summary, our FUSE observations have allowed us to derive a set of
stellar and wind parameters for young CSPN in the LMC that are
unhampered by the distance uncertainties that plague Galactic studies.
In addition to revealing the flux of the hot star (which is obscured
at longer wavelengths by the nebular flux), the far-UV observations
also revealed hot molecular hydrogen surrounding these young CSPN.

\acknowledgements

We thank John Hillier for his help with the CMFGEN code, as well as
for providing many useful comments.  We thank Stephan
McCandliss for making his \Htwo\ molecular data available.  We are
indebted to the members of the Opacity Project and Iron Project and to
Bob Kurucz for their continuing efforts to compute accurate atomic
data, without which, this project would not have been feasible.  The
SIMBAD database was used for literature searches.  This work has been
funded by NASA grants LTSA NAG-10364 and NAG 5-9219
(NRA-99-01-LTSA-029).  The HST data presented in this paper were
obtained from the Multimission Archive at the Space Telescope Science
Institute (MAST). STScI is operated by the Association of Universities
for Research in Astronomy, Inc., under NASA contract NAS5-26555.

\appendix

\section{APPENDIX: MODEL ATOMS}\label{sec:atomic}

Ions and the number of levels and superlevels included in the model
calculations are listed in Table~\ref{tab:ion_tab}.  The atomic data
come from a variety of sources, with the Opacity Project
\citep{seaton:87,opacity:95,opacity:97}, the Iron Project
\citep{pradhan:96,hummer:93}, \citet{kurucz:95}\footnote{See
http://cfa-www.harvard.edu/amdata/ampdata/amdata.shtml} and the Atomic
Spectra Database at NIST Physical Laboratory being the principal
sources.  Much of the Kurucz atomic data were obtained directly from
CfA \citep{kurucz:88,kurucz:02}.  Individual sources of atomic data
include the following: \citet{zhang:97}, \citet{bautista:97},
\citet{becker:95b}, \citet{butler:93}, \citet{fuhr:88},
\citet{luo:89a}, \citet{luo:89b}, Mendoza (1983, 1995, private
communication), \citet{mendoza:95},
\citet{nussbaumer:83,nussbaumer:84}, \citet{peach:88}, Storey (1988,
private communication), \citet{tully:90}, and
\citet{wiese:66,wiese:69}.  Unpublished data taken from the Opacity
Project include: \FeVI\ data (Butler, K.) and \FeVIII\ data (Saraph and
Storey).
% andC. Mendoza (\FeIX, \FeX).

%-----------------------------------------------------------------------

\newpage

%\bibliographystyle{apj}
%\bibliography{astro_refs}

%---------------------figures-------------------------------------------
\clearpage

\clearpage

\clearpage

\clearpage

\clearpage

\clearpage

\clearpage

\clearpage

\clearpage

\clearpage

\clearpage

\clearpage

\clearpage

\clearpage

\clearpage

\clearpage
%---------------------tables--------------------------------------------
%\newpage

\input{tab1}
\input{tab2}
\input{tab3}

\input{tab4}
\input{tab5}
\input{tab6}
\input{tab7}

\input{tab8}
\input{tab9}
\input{tab10}

\input{tab11}

\end{document}

%% file: tab1.tex
\begin{deluxetable}{lcclccccccc}
\rotate
\tabletypesize{\scriptsize}
\tablecolumns{10}
\tablewidth{0pc}
\tablecaption{FUSE B001 program stars, and FUSE detector segments utilized\label{tab:fuseobs}}
\tablehead{
\colhead{Name} & \colhead{R.A.} & \colhead{Dec} & 
\colhead{Archive} & \colhead{Date(s)} & \colhead{$t_{exp}$} & 
\multicolumn{4}{c}{Wavelength Range (\AA)}& \colhead{Background\tablenotemark{a}} \\
\colhead{} &\colhead{(2000)}&\colhead{(2000)}& 
\colhead{Name}    &\colhead{}      &  \colhead{(ks)} &
\colhead{SiC1}  &\colhead{SiC2} &\colhead{LiF1} &\colhead{LiF2}& \colhead{Model}
}
\startdata
SMP LMC~2 & 04 40 56.75& -67 48 02.70 & B0010201 & 08/20/01 & 9.3 &
920--993 & 920-995 & 995--1081,1095--1136 & 995--1075,1087--1183 & s\\
SMP LMC~23 & 05 06 09.43& -67 45 26.90 & B0010302 & 10/25/01 & 6.7 &
920--1000 & 920--1000 & 1000--1081 & 1000-1075,1087--1183 & d \\
SMP LMC~35 & 05 10 49.97 & -65 29 30.70 & B0010401-2 & 10/30-31/01  & 35.7
& - & 916--1006 & - & 1006--1074,1093--1183 & d\\
SMP LMC~61 & 05 24 35.97 & -73 40 39.68 & B0010501-7 & 4/17/02-5/23/02
&49.2 & - & 916--1000 & 1000--1075 & 1090--1183 & s\\
SMP LMC~62 & 05 24 55.16 & -71 32 56.32 & B0010601 & 3/20/02 & 11.3 &
920--1000 & 920--1000 & 1000-1081 & 1000-1075,1087--1183 & d\\
SMP LMC~67 & 05 29 15.75 & -67 32 47.58 & B0010701 & 11/14/01 & 38.4 &
- & 920--1000 & 1000--1080 & 1088--1183 & d\\
SMP LMC~85 & 05 40 30.87 & -66 17 37.53 & B0010801-2 & 10/10/01-02/14/02
&16.2 & - & 920--1004 & - & 1004--1082,1086--1183 & s\\
\enddata
%\tablecomments{}
\tablenotetext{a}{ ``d'' denotes calibration with default pipeline
  background subtraction, ``s'' with scattered light background model
  (see text).}
\end{deluxetable}

%% file: tab2.tex
\begin{table}
\caption{Archive UV data sets (FOS and STIS)}\label{tab:uvobs}
\begin{tabular}{lccccc}
\hline
Name   & Instrument & Archive & Date & Grating & $\lambda$  \\
       &            & Name    &      &         & (\AA) \\
\hline
SMP LMC~2  & FOS & Y1C10503T & 05/09/93 & G130H & 1182-1600\\
       & FOS & Y1C10504T & 05/09/93 & G190H & 1600-2330\\
       & FOS & Y1C10505T & 05/09/93 & PRISM & 2330-6015\\
SMP LMC~23 & FOS & Y2Y00104T & 10/21/95 & G130H & 1181-1520\\
       & FOS & Y2Y00104T & 10/21/95 & G130H & 1557-1575\\
       & FOS & Y2Y00105T & 10/21/95 & G190H & 1575-2250\\
       & FOS & Y2Y00106T & 10/21/95 & G270H & 2250-3280\\
       & FOS & Y2Y00107T & 10/21/95 & G400H & 3280-4822\\
SMP LMC~35 & FOS & Y1C10803T & 05/04/93 & G130H & 1182-1600\\
       & FOS & Y1C10804T & 05/04/93 & G190H & 1600-2300\\
       & FOS & Y1C10805T & 05/04/93 & PRISM & 2300-6015\\
SMP LMC~61 & STIS & O57N02010 & 01/07/99 & G140L & 1182-1720 \\
       & STIS & O57N02040 & 01/07/99 & G230L & 1720-3150 \\
       & STIS & O57N03010 & 01/07/99 & G430L & 3150-5200 \\
SMP LMC~62 & FOS & Y2Y00404T & 12/12/95 & G130H & 1182-1602 \\
       & FOS & Y2Y00405T & 12/12/95 & G190H & 1602-2300 \\
       & FOS & Y2Y00406T & 12/12/95 & G270H & 2300-3280 \\
       & FOS & Y2Y00407T & 12/12/95 & G400H & 3280-4822 \\
SMP LMC~67 & FOS & Y2N30402T & 03/20/95 & G130H & 1182-1600\\
       & FOS & Y2N30403T & 03/20/95 & G190H & 1600-2250\\
       & FOS & Y2N30405T & 03/20/95 & PRISM & 2250-5959 \\
SMP LMC~85 & FOS & Y17V0106T &  01/06/93 & G130H & 1182-1600\\
       & FOS & Y17V0104T &  01/06/93 & G190H & 1600-2250\\
       & FOS & Y17V0105T &  01/06/93 & G270H & 2250-5959\\
\hline
\end{tabular}
\end{table}

%% file: tab3.tex
\begin{table}
\caption{Line classifications and \vedge\ measurements}\label{tab:lines}
\scriptsize
\begin{tabular}{cccccccccc}
\hline
Ion     & \SVI    &\CIII& \OVI    & \CIII & \NV  & \OV  & \SiIV & \CIV    & \HeII \\
\hline
\singlet (\AA)& 933.4 & 977.0 & 1031.9 & 1174.9 & 1238.4 & 1371.3 & 1393.7  & 1548.2 & 1640.4 \\ 
%        & 944.5 &       & 1037.6  &        & 1242.8 &        & 1402.8  & 1550.8 &  \\ 
%IP (eV)     & 88      & 48  & 138     & 48    & 98  & 114  &  45   &   64    & 54 \\
\hline
SMP LMC~61  & ? & ps  & ps &1350$\pm$150& ?   &1300$\pm$200& - &1700$\pm$300&1200$\pm$200 \\
SMP LMC~2   & - &800$\pm$100&1050$\pm$150&700$\pm$100& pw   & - &700$\pm$100&1150$\pm$150&-\\
SMP LMC~62  & -       & a  &700$\pm$100& a & ? & - & neb & neb & neb \\
SMP LMC~23  &1350$\pm$200&350$\pm$100&1000$\pm$200 &650$\pm$100&1300$\pm$200& a & pw & 1100$\pm$100       & -  \\
SMP LMC~67  &1200$\pm$200&400$\pm$100&1300$\pm$200&a&1250$\pm$250& - & - &1100$\pm$100 & -\\
SMP LMC~85  & ? &1100$\pm$100&550$\pm$150& pw & ? & - & pw &850$\pm$200$^\dagger$& - \\ 
SMP LMC~35  & - & neb & 800$\pm$100 & neb & ? & - & pw & neb & neb  \\
\hline
\end{tabular}
\begin{minipage}{\textwidth}
\begin{trivlist}
\item Numbers are \vedge\ measurements of strong
  P-Cygni profiles in \kms.
\item ``-'': not present; ps: strong P-Cygni (but blue edge too
  unreliable for \vedge\ measurement); pw: weak P-Cygni; a: absorption; neb: nebular emisison; ?: presence questionable
\end{trivlist}
\end{minipage}
\end{table}

%% file: tab4.tex
\begin{deluxetable}{ccc}
\tabletypesize{\scriptsize}
\tablecolumns{2}
\tablewidth{0pc}
\tablecaption{SMP LMC~62 FUV and UV emission lines \label{tab:lmc62uv}}
\tablehead{
\colhead{$\lambda_{ID}$} & \colhead{Ion} & \colhead{Flux} \\
\colhead{(\AA)} &\colhead{} & \colhead{(\flam)}
}
\startdata
1032.43  &  \OVI  &  2.52E-14  \\
1037.97  &  \OVI  &  1.64E-14  \\
1239.67  &  \NV  &  6.56E-14  \\
1243.67  &  \NV  &  3.13E-14  \\
1394.56  &  \SiIV  &  2.21E-14  \\
1401.25  &  \SiIV b  &  9.25E-14  \\
1405.30  &  \OIV b  &  5.71E-14  \\
1483.79  &  \NIV]  &  4.72E-14  \\
1487.79  &  \NIV]  &  4.68E-14  \\
1548.74  &  \CIV  &  2.25E-13  \\
1552.19  &  \CIV  &  1.26E-13  \\
1575.32  &  [\NV]  &  6.86E-15  \\
1602.85  &  [\NeIV]  &  2.18E-14  \\
1642.02  &  \HeII  &  4.91E-13  \\
1662.11  &  \OIII]  &  4.15E-14  \\
1667.45  &  \OIII]  &  9.43E-14  \\
1752.94  &  \NIII]b  &  5.68E-14  \\
1883.94  &  \SiIII]  &  2.35E-14  \\
1893.18  &  \SiIII]  &  3.00E-14  \\
1908.77  &  \CIII  &  3.38E-13  \\
2325.86  &  [\OIII]b  &  4.71E-14  \\
2424.54  &  [\NeIV]  &  1.98E-13  \\
2471.70  &  \OII]  &  3.51E-14  \\
2513.07  &  \HeII  &  1.29E-14  \\
2735.15  &  \HeII  &  1.97E-14  \\
2784.57  &  \ArV /\MgV  &  2.39E-14  \\
2838.13  &  \OIII?  &  4.23E-14  \\
2859.71  &  \ArIV  &  8.24E-15  \\
2870.00  &  \ArIV  &  2.81E-15  \\
3025.79  &  \OIII  &  9.61E-15  \\
3048.92  &  \OIII  &  2.12E-14  \\
3123.71  &  ?  &  1.36E-14  \\
3135.01  &  \OIII  &  1.54E-13  \\
3189.81  &  \HeI  &  1.06E-14  \\
3204.80  &  \HeII  &  4.20E-14  \\
3315.67  &  ?  &  1.58E-14  \\
3348.00  &  [\NeV]  &  1.48E-13  \\
3430.16  &  [\NeV]  &  3.55E-13  \\
3447.27  &  ?  &  4.32E-14  \\
3731.31  &  [\OII]  &  1.80E-13  \\
3801.36  &  ?  &  1.88E-14  \\
3840.78  &  ?  &  3.01E-14  \\
3872.40  &  [\NeIII]  &  3.67E-13  \\
3892.64  &  ?  &  8.11E-14  \\
3971.43  &  [\NeIII]  &  1.81E-13  \\
4072.75  &  [\SII]  &  1.42E-14  \\
4106.68  &  \Hdelta  &  1.10E-13  \\
4344.45  &  \Hgamma  &  2.02E-13  \\
4367.73  &  [\OIII]  &  9.08E-14  \\
4475.50  &  \HeI  &  1.17E-14  \\
4690.59  &  \HeII  &  9.29E-14  \\
\enddata
\tablenotetext{b}{Line is blended.}
\tablenotetext{?}{Unidentified feature.}
\end{deluxetable}

%% file: tab5.tex
\begin{deluxetable}{cccccccccc}
\tablecolumns{10}
\tablewidth{0pc}
\tablecaption{\HI\ and \Htwo\ parameters{\label{tab:hydrogen}}}
\tablehead{
\colhead{} & \multicolumn{2}{c}{$\HI^{IS + circ}$} & \colhead{} & \multicolumn{2}{c}{$\Htwo^{IS}$} &
\colhead{} & \multicolumn{2}{c}{$\Htwo^{circ}$} & \colhead{}\\
\cline{2-3}\cline{5-6}\cline{8-9}
\colhead{Star} & \colhead{\logN} & \colhead{$T$} & \colhead{} &
\colhead{\logN} & \colhead{$T$} & \colhead{} & \colhead{\logN} &
\colhead{$T$} & \colhead{\EBMV (\HI)}\\
\colhead{} &\colhead{(cm$^{-2}$)} &\colhead{(K)} & \colhead{} &
\colhead{(cm$^{-2}$)} &\colhead{(K)} & \colhead{} & \colhead{(cm$^{-2}$)}
&\colhead{(K)} & \colhead{(mag)} \\
}
\startdata
SMP LMC~62 & $21.70^{+0.17}_{-0.30}$ & 80 && $16.70^{+0.30}_{-0.40}$ &
80 &&$16.70^{+0.30}_{-0.40}$  & $3000\pm1000$ & 1.0$^{+0.5}_{-0.5}$ \\
SMP LMC~23 & $20.70^{+0.30}_{-0.30}$ & 80 && 17.70$^{+0.30}_{-0.70}$ &
80 && 16.70$^{+0.30}_{-0.70}$ & $3000\pm1000$ & 0.10$^{+0.10}_{-0.05}$\\
SMP LMC~67 & 21.00$^{+0.40}_{-0.70}$ & 80 && 18.40$^{+0.30}_{-0.40}$ & 80 && 17.00$^{+0.40}_{-0.30}$ & $3000\pm1000$& 0.20$^{+0.30}_{-0.16}$\\
SMP LMC~61 & $21.40^{+0.30}_{-0.40}$ & 80 && $20.00^{+0.40}_{-0.70}$ & 80 && $17.00^{+0.70}_{-0.30}$ & $2000\pm1000$& 0.5$^{+0.5}_{-0.3}$\\
SMP LMC~85 & 21.40$^{+1.00}_{-1.00}$ & 80 && $\leq$20.00 & 80 && 17.00$^{+0.70}_{-0.30}$ & $3000\pm1000$& 0.5$^{+4.5}_{-0.45}$ \\
SMP LMC~2  & 21.00$^{+0.40}_{-0.60}$ & 80 && 19.70$^{+0.30}_{-0.40}$ &
80 && 17.00$^{+0.30}_{-0.30}$ & $3000\pm1000$ & 0.20$^{+0.30}_{-0.15}$\\
SMP LMC~35 & 21.0 & 80 && 18.7$^{+0.30}_{-0.70}$ & 80 && 16.7$^{+0.30}_{-0.70}$ & 4000& -\\
\hline
\enddata
\end{deluxetable}

%% file: tab6.tex
\begin{deluxetable}{lcccccccccc}
\rotate
\tabletypesize{\footnotesize}
\tablecolumns{12}
\tablewidth{0pc}
\tablecaption{Nebular Parameters \label{tab:neb}}
\tablehead{
\colhead{Name} & 
\colhead{$\theta^{aa}$} &
\colhead{$r_{neb}^{bb}$} &
\colhead{$v_{exp}(\rm{OIII})^{a}$} &
\colhead{$\tau_{dyn}^{cc}$} &
\colhead{$n_{e}([\rm{OII}])$} & 
\colhead{\Telec(O$^{+2}$)$^{b}$} & 
\colhead{$\log(F_{H\beta})^{a}$} &
\colhead{\EBMV(c$_{H\beta}$)$^{b}$} & 
\colhead{He/H$^c$} & 
\colhead{He$^{2+}$/He$^+$} \\
\colhead{} & 
\colhead{(\arcsec)} & 
\colhead{(pc)} & 
\colhead{(\kms)} & 
\colhead{(kyr)} &
\colhead{(10$^3$ cm$^{-3}$)} & 
\colhead{(kK)} & 
\colhead{(ergs cm$^{-2}$ s$^{-1}$)} & 
\colhead{(mag)} & 
\colhead{} & 
\colhead{} 
}
\startdata
SMP LMC~61 & 0.465 & 0.057$^{d}$ & 29.3 & 1.9 &  40.0$^{a}$ & 11.5 &
-12.48 (\textbf{-12.55}) &0.13& 0.140 & 0.0$^{e}$ \\
SMP LMC~23 & 0.705 & 0.086$^f$ & 21.6 & 3.9 & 5.9$^{a}$ & 11.2 &
-12.68 (\textbf{-12.75}) &0.04 & 0.095 & 0.0$^{e}$ \\
SMP LMC~67 & 0.648 & 0.079$^{d}$ & 27.9 & 2.8 & 3.7$^{a}$ & 12.2 &
-12.81 (\textbf{-12.63}) &0.10 & 0.139 & 0.157$^{e}$\\
SMP LMC~62 & 0.588 & 0.068$^{d}$ & 34.6 & 1.9 & 6.2$^{a}$ & 15.7
& -12.31 (\textbf{-12.43}) &0.14 & 0.088 & 0.333$^{e}$ \\
SMP LMC~85 & $<0.163$ & $<0.02^{f}$ & 11.3 & $<1.7$ & ``high''$^{a}$ (\textbf{40})&
11.1  & -12.42 (\textbf{-11.91}) &0.17 & 0.140 & -\\
SMP LMC~2  & 0.542 & 0.066$^{f}$ & 9.9 & 6.5 & - (\textbf{5}) & 10.5 &
-13.18 (\textbf{-12.90}) &0.07&  0.128 & -\\
SMP LMC~35 & 1.88 & 0.231$^{d}$ & 41.3$^{dd}$ & 5.5 & 1.6$^{ee}$ &
13.3  &-12.81 (\textbf{-12.96}) & 0.00 & 0.063 & 0.254$^{e}$\\
\enddata
\tablenotetext{aa}{Calculated from $r_{neb}$  using LMC distance of 50.6~kpc}
\tablenotetext{bb}{From their ``D(edge)''}
\tablenotetext{cc}{$\tau_{dyn}=r_{neb}/v_{exp}$}
\tablenotetext{dd}{Line fit with two components}
\tablenotetext{ee}{From [\SII] \citep{meatheringham:91b}}
\tablerefs{
(a): \citet{dopita:88}
(b): \citet{meatheringham:91a,meatheringham:91b}
(c): \citet{dopita:91a,dopita:91b}
(d): \citet{vassiliadis:98}
(e): \citet{monk:88}
(f): \citet{dopita:96}
}
\end{deluxetable}

%% file: tab7.tex
\begin{table}
\scriptsize
\caption{Abundances of grid models}\label{tab:abund}
\begin{tabular}{cccccccccc}
\hline
Grid    & \XH & \XHe & \XC & \XN & \XO & \XSi & \XP & \XS & \XFe \\
\hline
%He-rich & - & 0.54 & 0.37 & 0.01 & 0.08 & \Xlmc & \Xlmc & \Xlmc &\Xlmc \\
H-deficient & - & 0.54 & 0.37 & 0.01 & 0.08 & 2.80\E{-4} & 2.45\E{-6} & 1.53\E{-4} & 5.44\E{-4}\\
%H-rich & \Xlmc & \Xlmc & \Xlmc & \Xlmc & \Xlmc & \Xlmc & \Xlmc & \Xlmc& \Xlmc\\
H-rich & 0.71 & 0.28 & 1.22\E{-3} & 4.40\E{-4} & 3.82\E{-3} &
2.80\E{-4} & 2.45\E{-6} & 1.53\E{-4} & 5.44\E{-4}\\
\hline
\end{tabular}
\end{table}

%% file: tab8.tex
\begin{deluxetable}{lccccccccccc}
\rotate
\tabletypesize{\footnotesize}
\tablecolumns{11}
\tablewidth{0pc}
\tablecaption{Derived Stellar Parameters \label{tab:mod_param}}
\tablehead{
\colhead{Name} & 
\colhead{Model} & 
\colhead{\Teff} & 
\colhead{$\log{L}$} & 
\colhead{\Rstar} & 
\colhead{$\log{\Mdot_{s}^{a}}$} & 
\colhead{$\log{\Mdot_{c}^{b}}$} & 
\colhead{\vinf} &
\colhead{$M^{c}$} & 
\colhead{\logg} & 
\colhead{$\Gamma$} & 
\colhead{\EBMV} \\
\colhead{} & 
\colhead{Abundance} & 
\colhead{(kK)} & 
\colhead{(\Lsun)}& 
\colhead{(\Rsun)} & 
\colhead{(\Msunyr)} & 
\colhead{(\Msunyr)} & 
\colhead{(\kms)} & 
\colhead{(\Msun)} & 
\colhead{(cgs)}& 
\colhead{} & 
\colhead{(mag)}  
}
\startdata
SMP LMC~61 & H-poor & $70\pm5$ & $3.62^{+0.06}_{-0.11}$ &$0.44\pm0.02$ &
$-6.33\pm0.33$ & $-6.83\pm0.33$& $1300\pm200$ & $0.64$ & $4.96$ & 0.20 &0.05\\
SMP LMC~23 & H-rich & $60\pm5$ & $3.77\pm0.07$
&$0.72^{+0.80}_{-0.05}$&$-7.92\pm0.35$ &$-7.42\pm0.35$ & $1100\pm100$ &
$0.66$ &$4.54$ & 0.28 & 0.04 \\
SMP LMC~67 & H-rich & $55\pm5$ & $3.70\pm0.10$ & $0.80^{+0.10}_{-0.05}$ &
$-7.35\pm0.35$ & $-7.85\pm0.35$& $1000\pm200$ & $0.65$ &$4.44$&0.24 & 0.09\\
SMP LMC~62 & H-rich & $45\pm5$ & $3.73^{+0.11}_{-0.13}$ &
$1.21\pm0.11$&$-7.78^{+0.75}_{-0.36}$ &$-8.28^{+0.75}_{-0.36}$ & $1000\pm300$ & $0.65$
&$4.30$& 0.25 & 0.09  \\
SMP LMC~85 &  H-rich & $40\pm2$ & $3.41\pm0.05$ & $1.06\pm0.05$ &
$-7.3\pm0.3$  &$-7.8\pm0.3$ &
$700\pm100$  & $0.57$ &$4.14$& 0.14 & 0.13 \\
SMP LMC~2  &  H-rich & $38\pm2$ & $3.42\pm0.06$ & $1.18\pm0.05$ &
$-7.55^{+0.3}_{-0.4}$ & $-8.05^{+0.3}_{-0.4}$ & $700\pm100$ & $0.56$ &$4.04$&
0.15 & 0.07  \\
SMP LMC~35 &  H-rich & 60 & 3.03 & 0.3 & -7.97&  -8.47& 1000 &$0.55$ & 5.20& -
& 0.03 \\
\enddata
\tablenotetext{a}{$\Mdot_s$ denotes smooth (unclumped) mass-loss rate
  corresponding to a filing factor of $f=1.0$.}
\tablenotetext{b}{$\Mdot_c$ denotes clumped mass-loss rate
  corresponding to a filing factor of $f=0.1$.}
\tablenotetext{c}{From evolutionary tracks of \citet{vassiliadis:94}.}
\end{deluxetable}

%% file: tab9.tex
\begin{table}
\scriptsize
\caption{Comparison Parameters}\label{tab:photo}
\begin{tabular}{lccccccccc}
\hline
Name & $T_{eff}^{photo}$ & $\log{\Mdot}^{\dagger}$&
$\log{\Mdot}^{\ddagger}$&  $\tau_{dyn}$ & $\tau_{evol}$ &
$v_{LRS}^{a}$ &$\eta$ & $\Pi$\\
 & (kK) & (\Msunyr) & (\Msunyr)  &(kyr) & (kyr) & (\kms) &  &  ($\Msunyr \kms \Rsun^{0.5}$)\\
\hline
SMP LMC~61 & 59 & -8.20 & -7.99 & 1.9 & 2-5 & 178.4 & $7^{+13}_{-4}$& -3.39 \\
SMP LMC~23 & 65 & -7.77 & -7.51 & 3.9 & 2-3 & 267.1 & $0.5\pm0.4$& -4.49 &\\
SMP LMC~67 & 46 & -7.74 & -7.48 & 2.8 & 1-3 & 274.1 & $0.7\pm0.5$ & -4.40\\
SMP LMC~62 &127 & -7.79 & -7.53  & 1.9 & 1-2 & 223.6 &$0.6\pm0.5$ & -4.74\\
SMP LMC~85 & 45 & -8.16 & -7.91  & $<$1.7 & 1-3 & 217.0 & \textbf{$1.0\pm0.8$}& -4.38\\
SMP LMC~2  & 39 & -8.21 & -7.95 & 6.5 & 1-3 & 248.2 & $0.5\pm0.4$ & -4.67\\
SMP LMC~35 & 118 &- & - & 5.5 & - & 297.0 & 0.5 & -\\
\hline
\end{tabular}
\begin{minipage}{\textwidth}
\begin{trivlist}
\item $\dagger$ Using prescription of \citet{vink:01} with $z=0.4\Zsun$
\item $\ddagger$ Using prescription of \citet{vink:01} with $z=0.8\Zsun$
\item $a$ From \citet{dopita:88}
\end{trivlist}
\end{minipage}
\end{table}

%% file: tab10.tex
\begin{table}
\scriptsize
\caption{SMP LMC~61 abundances}\label{tab:lmc61abund}
\begin{tabular}{ccccccccc}
\hline
\XH & \XHe & \XC & \XN & \XO & \XSi & \XP & \XS & \XFe \\
\hline
%He-rich & - & 0.54 & 0.37 & 0.01 & 0.08 & \Xlmc & \Xlmc & \Xlmc &\Xlmc \\
%He-rich & - & 0.54 & 0.37 & 0.01 & 0.08 & 2.80\E{-4} & 2.45\E{-6} & 1.53\E{-4} & 5.44\E{-4}\\
0.00 & 0.54 & 0.37 & $<0.01$ & 0.08 & 2.80\E{-4} & 2.45\E{-6} & 1.53\E{-4} & 5.44\E{-5}\\
\hline
\end{tabular}
\end{table}

%% file: tab11.tex
\begin{table}
\caption{Levels and superlevels for model ions}\label{tab:ion_tab}
\footnotesize
\begin{tabular}{lcccccccccc}
\hline
Element & I & II & III & IV & V & VI & VII & VIII & IX \\
\hline
H  & 20/30 & 1/1 & & & & & & & & \\
He & 40/45 & 22/30 & 1/1\\
C  & & 14/14 & 30/54 & 13/18 & 1/1\\
N  & & & 34/70 & 29/53 & 13/21 & 1/1\\
O  & & 24/24 & 25/45 & 29/48 & 41/78 & 13/19 & 1/1\\
Si & & & 20/34 & 22/33 & 1/1\\
P  & & & & 36/178 & 16/62 & 1/1\\
S  & & & 13/28 & 51/142 & 31/98 & 28/58 & 1/1\\
Fe & & & & 51/294 & 47/191 & 44/433 & 41/254 & 53/324& 1/1\\
\hline
\end{tabular}
\end{table}